\documentclass[opre,nonblindrev]{informs3-arXiv} 

\DoubleSpacedXI 

\usepackage{natbib}
 \bibpunct[, ]{(}{)}{,}{a}{}{,}%
\usepackage{mathrsfs}
\usepackage{threeparttable}
\usepackage{multirow}
\usepackage{diagbox}
\usepackage{booktabs}
\usepackage{textcomp}
\newcommand{\tabincell}[2]{
\begin{tabular}{@{}#1@{}}#2\end{tabular}
}
\usepackage{color}
\usepackage{subfigure}
\usepackage{makecell}
\usepackage{caption}
\usepackage{listings}
\usepackage{tikz}
\usetikzlibrary{decorations.pathreplacing}

\usepackage{slashbox}

\TheoremsNumberedThrough     
\ECRepeatTheorems

\EquationsNumberedThrough    

\MANUSCRIPTNO{} 

\begin{document}


\RUNAUTHOR{~}

\RUNTITLE{Option Pricing under MS-SVCJ Model}

\TITLE{Option Pricing Under a Discrete-Time Markov Switching Stochastic Volatility with Co-Jump Model}

\ARTICLEAUTHORS{%
\AUTHOR{Michael C. Fu}
\AFF{Smith School of Business \& Institute for Systems Research, University of Maryland, College Park, MD 20742,
\EMAIL{mfu@umd.edu}} 
\AUTHOR{Bingqing Li}
\AFF{School of Finance, Nankai University, 300350 Tianjin, China, \EMAIL{libq@nankai.edu.cn}}
\AUTHOR{Rongwen Wu} \AFF{A Financial Company in the US, \EMAIL{rongwen\_wu@hotmail.com}}
\AUTHOR{Tianqi Zhang}
\AFF{School of Finance, Nankai University, 300350 Tianjin, China, \EMAIL{zhangtq@mail.nankai.edu.cn}}
} 

\ABSTRACT{%

We consider option pricing using a discrete-time Markov switching stochastic volatility with co-jump model, which can model volatility clustering and varying mean-reversion speeds of volatility. For pricing European options, we develop a computationally efficient method for obtaining the probability distribution of average integrated variance (AIV), which is key to option pricing under stochastic-volatility-type models. Building upon the efficiency of the European option pricing approach, we are able to price an American-style option, by converting its pricing into the pricing of a portfolio of European options. Our work also provides constructive guidance for analyzing derivatives based on variance, e.g., the variance swap. Numerical results indicate our methods can be implemented very efficiently and accurately.
}%



\maketitle

%


\section{Introduction}
The stochastic volatility with co-jump (SVCJ) model introduced by \cite{RN356} is commonly used to model the price dynamics for an underlying financial asset, because it is able to capture leptokurtic, skewness, and volatility clustering observed in real-world data. The SVCJ model simultaneously considers stochastic volatility, jumps in return, and jumps in volatility, generalizing the jump-diffusion model in \cite{RN342}, the stochastic volatility model in \cite{RN369}, and the stochastic volatility/jump-diffusion model in \cite{RN334}, which are all special cases of SVCJ. Empirical evidence supporting the presence and importance of stochastic volatility with jumps in both return and volatility is documented in \cite{RN356}, and surveys of critical developments in the SVCJ model can be found in \cite{RN614}, \cite{RN336}, \cite{RN822}, \cite{RN825}, \cite{RN265}, \cite{RN298} and references therein. SVCJ models primarily price options using Fourier transform methods.

However, \cite{RN703} and \cite{RN235} point out that the SVCJ model cannot sufficiently capture volatility clustering, especially for scenarios with persistently high levels of volatility such as the period from Feb 28 to May 15, 2020, since the CIR process used in the SVCJ model assumes rapid mean reversion of volatility. \cite{RN652} observes that the volatility process in the SVCJ model spends much time close to zero, so once an extreme value in volatility is reached, the preference for moving back to zero volatility leads to an excessively rapid mean-reversion speed. \cite{RN614} also empirically observed slower mean-reversion speed improving performance of their model in the out-of-sample period. Here we also refer to the SVCJ model with CIR process as the classical SVCJ model.\par

To address some of the shortcomings of the classical SVCJ model, we propose a discrete-time Markov switching stochastic volatility with co-jump (MS-SVCJ) model where the volatility is composed of a Markov switching (MS) process and a jump process. \cite{RN343}, \cite{RN767}, and \cite{RN232} pointed out that MS processes can model the persistence of volatility. For example, \cite{RN343} stressed that, ``by choosing the parameters appropriately, we can model different levels of persistence of the volatility process in the high and low states.'' Moreover, \cite{RN767} stated that ``Markov switching models can generate a wide range of coefficients of skewness, kurtosis and serial correlation even when based on a very small number of underlying states.'' Similar arguments are made in related literature, e.g., \cite{RN357}, \cite{RN779}, \cite{RN614}, \cite{RN819}, \cite{RN788}, \cite{RN773}, \cite{RN628}, \cite{RN761}.

In addition, since the MS process can distributionally approximate any diffusion stochastic process, we can approximate the CIR process by adjusting the parameters of the MS process. Detailed approximation implementations can be found in \cite{RN388}, \cite{RN3932} and \cite{RN541}. The proposed model can distributionally approximate the classic SVCJ model, so it is more robust and flexible than the classical SVCJ model. In short, the proposed model overcomes some limitations of the classical SVCJ model while retaining its advantages. \par

For pricing European options under the proposed model, we consider the method based on average integrated variance (AIV) developed by \cite{RN361}, in which the option price under the stochastic volatility model is expressed as the expectation of the Black-Scholes formula with variance replaced by AIV. Although this provides a formal solution for the option price, the probability distribution of AIV is generally difficult to obtain, which makes the practical application of this method challenging. For our MS-SVCJ model, we face the same challenge. \par

We consider a discrete-time MS volatility process with finite state space, so there are a finite number of sample paths of volatility, which means the value space of AIV is also finite. Thus, theoretically we can find the probability distribution of AIV by enumerating all the sample paths of volatility, but such enumeration is generally not computationally feasible, so we propose the recursive recombination (RR) algorithm to efficiently compute the probability distribution of AIV. Then we derive a pricing formula that leads to an analytical solution for European options under our proposed model. Numerical experiments demonstrate the effectiveness of the RR algorithm. \par

Our work extends the existing literature on European option pricing in several ways. First, without the jumps in volatility, the proposed model becomes the MS-SVJ model, so our analytical solution for European options applies to the MS-SVJ model with general jump size distribution. Second, removing jumps from both the volatility and asset price, the proposed model reduces to the MS-SV model, so using our analytical solution for the MS-SV model avoids having to solve a set of intractable ordinary differential equations when pricing options, as in \cite{RN343}, \cite{RN232}, and \cite{RN374}, or using numerical inverse Fourier transform methods, as in \cite{RN409}, \cite{RN590}, and \cite{RN469}. \par

We also consider the pricing of American-style derivatives (see \cite{RN3933}), applying the approach proposed by \cite{RN1247} by leveraging our analytical solutions for pricing European options. In the approach, the pricing of an American-style option can be converted into the pricing of a basket of European options. We compare our results with the least squares Monte Carlo simulation approach by \cite{RN3931}. For more discussion of the application of Monte Carlo simulation approaches to American-style derivatives, refer to \cite{RN3935,RN3934} or \cite{RN3936}. Our numerical examples further illustrate the effectiveness of our approach. \par

In sum, our work contributes to the option pricing research literature as follows:
\begin{itemize}
  \item We provide an analytical pricing formula for European options under a discrete-time MS-SVCJ model that is more robust and flexible than the classical SVCJ model and can explain volatility clustering for high levels of volatility.
  \item We develop an efficient algorithm to obtain the probability distribution for AIV, which can also be applied to other related volatility derivatives, e.g., the variance swap.
  \item Our analytical solution for European option prices applies to several well-known models in the literature, including the MS-SVJ model with general jump size distribution. For the MS-SV model, our approach also has computational advantages over existing option pricing methods, by eliminating having to numerically solve a set of ordinary differential equations. 
  \item We price American-style options by leveraging the efficiency of our analytical pricing solution for European options.

\end{itemize}

The remainder of the paper is organized as follows. In Section 2, we construct the MS-SVCJ model and analyze the AIV probability measure. Section 3 develops the option valuation under the MS-SVCJ model. In Section 4, we present the RR algorithm and analyze its computational complexity. Numerical results for pricing both European options and American-style options are presented in Section 5. Section 6 concludes and discusses future research. \par

\section{Theoretical Framework}
\vspace*{-6pt}
In this section, we build the Markov switching stochastic volatility with co-jump (MS-SVCJ) model for the underlying asset and define the AIV probability measure.

\vspace*{-6pt}
\subsection{Model Setting}
\vspace*{-6pt}
Under the MS-SVCJ model, the underlying asset price $S_{t}$ is assumed to follow a jump-diffusion process, and the asset volatility is also stochastic. Specifically, the dynamics are specified by the following two equations under the risk-neutral probability measure $\mathbb{Q}$:
\begin{equation}\label{original}
  \begin{aligned}
  \frac{dS_{t}}{S_{t_{-}}}&=\left(r-\lambda \zeta\right)dt+\hat{\sigma}_{t}dB_{t}+\left(J_{t}-1\right)dN_{t}\\
  \hat{\sigma}_{t}^{2}&=\sigma_{t}^{2}+\sum_{i=1}^{N_{t}}f\left(J_{i},t,t_{i}\right)
  \end{aligned}
\end{equation}
where $\{B_{t}\}$ is standard Brownian motion; $\{N_{t}\}$ is a Poisson jump process with intensity $\lambda$; the proportional jumps $\{J_{t}\}$ are independent and identically distributed (i.i.d.) with $\zeta\equiv\mathbb{E}(J_{t}-1)$; $\{\sigma_{t}\}$ follows a discrete-time Markov switching (MS) process; $f(\cdot)$ which states the impact of jumps in asset price on variance is the proportional and exponentially attenuating (PEA) process; $\{B_{t}\}$, $\{N_{t}\}$, $\{J_{t}\}$ and $\{\sigma_{t}\}$ are mutually independent. We regard the risk-free interest rate $r$ as constant.\par

The main distinctive feature of the proposed model is that the variance process $\{\hat{\sigma}_{t}^{2}\}$ is composed of two components: the first explains the exogenous dynamics of variance, e.g., due to changes in the economy and company announcements,  and the second explains the endogenous movement in variance due to jumps in the asset, similar to \cite{RN370}. The proposed model assumes the first part follows an MS process and the second part follows the jump process related to jump in asset price. In what follows, we provide detailed specifications of the two processes.\par

We assume $\{\sigma_{t}\}$ follows a discrete-time Markov switching (MS) process with finite state space $\{u_{1}, u_{2},\cdots, u_{m}\}$ and constant time step $\tau$, and one-step transition probability matrix $P=[p_{ij}]_{m\times m}$, i.e.,
$
     p_{ij}=p(\sigma_{(k+1)\tau}=u_{j}|\sigma_{k\tau}=u_{i},\sigma_{(k-1)\tau},\cdots \cdots)
     =p(\sigma_{(k+1)\tau}=u_{j}|\sigma_{k\tau}=u_{i}).
$
The MS process has been shown to model reasonably well most of the stylized facts of volatility, volatility clustering and mean-reversion. See \cite{RN343}, \cite{RN576}, \cite{RN250}, \cite{RN199}, \cite{RN431}. Furthermore, compared with affine models for volatility, e.g. the square-root model, the MS process can better capture the clustering in different volatility levels and the varying mean-reversion speeds of volatility. \par
The second part $\sum\limits_{i=1}^{N_{t}}f\left(J_{i},t,t_{i}\right)$ takes into account the sudden movement in variance caused by jumps in the asset price. Here, we model $f\left(J_{i},t,t_{i}\right)$ as the product of two components, one representing the instantaneous shock size of variance due to the jump in asset return, and the other representing the dynamics of this shock over time. This dependence structure of jumps in asset price and volatility was proposed theoretically by \cite{RN330}, \cite{RN442}. \cite{RN373} empirically identified the effectiveness of this structure. Specifically, we give the following description of $f\left(J_{i},t,t_{i}\right)$.
\begin{definition}\label{def2}
 The $\{f\left(J_{i},t,t_{i}\right)\}$ is a proportional and exponentially attenuating (PEA) process if $f\left(J_{i},t,t_{i}\right)=f_{1}\left(J_{i}\right)f_{2}\left(t,t_{i}\right)$ for the $ith$ jump $J_{i}$ in asset price at $t_{i}$, where
\begin{align*}
f_{1}\left(J_{i}\right)&=b\ln^{2}\left(J_{i}\right), \\
f_{2}\left(t,t_{i}\right)&=
  e^{-\beta(t-t_{i})}, ~t_{i}<t\leq t_{i}+\Delta; ~~
  0, \mbox{~~otherwise}.
\end{align*}
\end{definition}

In the above representation, the term $f_{1}\left(J_{i}\right)$ is the shock size of variance caused by the jump in asset price, and the shock size is proportional to the square of log-jump in asset price with proportional coefficient $b$, which is consistent with the definition of variance expressed as the average of the quadratic function of the decentralized logarithm of the return. As a memory function, the term $f_{2}\left(t,t_{i}\right)$ indicates that the shock of the jump tails off exponentially with attenuating factor $\beta$ and duration $\Delta$, which is analogous to the CARMA kernel in \cite{RN237}. Following \cite{RN657}, we assume that the duration $\Delta$ is finite and fixed.

\vspace*{-6pt}
\subsection{Probability Measure}\label{subsection2}
\vspace*{-6pt}
\cite{RN361} show that the option price under the stochastic volatility model can be computed as the expectation of the Black-Scholes formula with variance replaced by average integrated variance (AIV). We derive a similar result in Section \ref{subsec2} when the diffusive innovation to the asset price process is independent of volatility. Under the proposed model, AIV during the interval $[0,T]$ can be expressed as
  \begin{equation}\label{defaiv}
      \mathscr{V}=\frac{1}{T}\int_{0}^{T}\hat{\sigma}_{t}^{2}dt
      =\frac{1}{T}\int_{0}^{T}\sigma_{t}^{2}dt+\frac{1}{T}\int_{0}^{T}\sum_{i}f(J_{i},t,t_{i})dt.
  \end{equation}
Thus, AIV can be expressed as a sum of two terms, one due to the MS process and the other due to the PEA process. Since AIV due to the MS process plays a key role, we first provide the following description.
We assume that $\tau$ is chosen such that $L={T}/{\tau}$ is integer-valued, so that $L$ is the total number of time steps. Since the information up to current time is available, we  also assume the initial state of $\{\sigma_{t}\}$ is known. The MS process $\{\sigma_{t}\}$ generates sample path $\omega$  during $[0,T]$. Since $\{\sigma_{t}\}$ is a piecewise constant process, we can represent the sample path $\omega$ as the following tuple form:
$
  \omega =\left[\sigma_{0},\sigma_{\tau},\sigma_{2\tau},\ldots,\sigma_{(L-1)\tau},\sigma_{L\tau}\right],
$
where $\sigma_{0}$ is fixed.

For notational simplicity, we write $\sigma_{k\tau}$ as $\sigma_{k}$. In the following, the MS process $\{\sigma_t\}$ will be rewritten as $\{\sigma_k\}$ to highlight its discretization, and henceforth, the above sample path $\omega$ will be expressed as
$
  \omega =\left[\sigma_{0},\sigma_{1},\sigma_{2},\ldots,\sigma_{L-1},\sigma_{L}\right],
$
with weight $|\omega|$ and probability $p(\omega)$ given by
$$
  \begin{aligned}
  |\omega| &=\frac{\sigma_{0}^{2}+\sigma_{1}^{2}+\sigma_{2}^{2}+\ldots+\sigma_{L-2}^{2}+\sigma_{L-1}^{2}}{L}, \\
  p(\omega) &=p_{\sigma_{0}\sigma_{1}}\cdot p_{\sigma_{1}\sigma_{2}}\cdots p_{\sigma_{L-2}\sigma_{L-1}}\cdot p_{\sigma_{L-1}\sigma_{L}}. \\
  \end{aligned}
$$
Here, we denote the set of all sample paths $\omega$ as $\Omega$, the sample path space for $\{\sigma_{k}\}$.

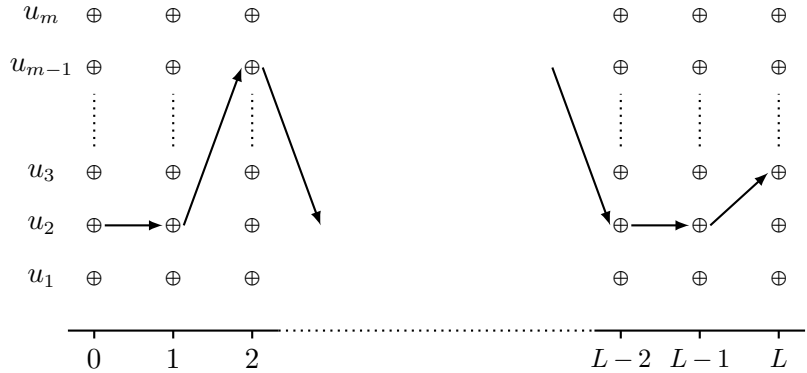
\begin{figure}[tb]
  \centering
\begin{tikzpicture}[scale=0.7]
\draw[thick] (0,0) -- (4,0);
\draw[dotted,thick] (4,0) -- (10,0);
\draw[thick] (10,0) -- (14,0);
\foreach \x/\y in{0.5/$0$,2/$1$,3.5/$2$,10.5/\small{$L-2$},12/\small{$L-1$},13.5/\small{$L$}} {\draw[thick] (\x,0) -- (\x,-0.15) node [below]{\y};}

\foreach \x/\y in{6/$u_{m}$,5/$u_{m-1}$,3/$u_{3}$,2/$u_{2}$,1/$u_{1}$} {\node at (-0.5,\x) {\y};}

\foreach \y in {0.5,2,3.5,10.5,12,13.5}{ \foreach \x in {1,2,3,5,6} {\node at (\y,\x) {\scalebox{0.5}[0.5]{$\bigoplus$}};} }

\foreach \x in {0.5,2,3.5,10.5,12,13.5} {\draw[dotted,thick] (\x,3.5) -- (\x,4.5);}

\draw[thick,-latex] (0.7,2) -- (1.8,2);
\draw[thick,-latex] (2.2,2) -- (3.3,5);
\draw[thick,-latex] (3.7,5) -- (4.8,2);
\draw[thick,-latex] (9.2,5) -- (10.3,2);
\draw[thick,-latex] (10.7,2) -- (11.8,2);
\draw[thick,-latex] (12.2,2) -- (13.3,3);
\end{tikzpicture}
\vspace*{-6pt}
\caption{Sample Path $\omega$: $\sigma_{0}=u_{2}$, $\sigma_{1}=u_{2}$, $\sigma_{2}=u_{m-1}$, \ldots, $\sigma_{L-2}=u_{2}$, $\sigma_{L-1}=u_{2}$, $\sigma_{L}=u_{3}$}
  \label{figure1}
\end{figure}

Figure $\ref{figure1}$ illustrates an example of sample path $\omega=[u_{2},u_{2},..., u_{m-1},\ldots,u_{2},u_{2},u_{3}]$.

AIV due to the MS process, which is the first term defined in Equation $(\ref{defaiv})$, is given by
\begin{equation*}
  V=\frac{1}{L}\sum_{k=1}^{L}\sigma_{k-1}^{2},
\end{equation*}
which is a random variable with probability distribution derived from $|\omega|$ and $p(\omega)$ with value space
$
  \Psi=\{v: v=|\omega|, \omega \in \Omega\},
$
and corresponding probability
$
p_{V}(v):=p(V=v)=
\sum\limits_{\omega \in \Omega :|\omega|=v}p(\omega), v \in \Psi.
$
Clearly, $|\Psi|\leq |\Omega|$, and in general, the number of possible values of $V$ is far less than the total number of sample paths of $\{\sigma_{k}\}$, and we have the following proposition (see Appendix \ref{numdis} for proof).
\begin{proposition}\label{pro2}
  $|\Psi| \leq \binom{L+m-2}{m-1}$.
\end{proposition}

\section{Option Valuation}
\vspace*{-6pt}

\subsection{European Option Pricing}\label{subsec2}
\vspace*{-6pt}
In this section, we price a European call option under the proposed model described by Equation $(\ref{original})$. A European put option can also be priced by the same method. First, we provide the following formal solution for the European call option price (see Appendix \ref{lemma} for proof).
\begin{lemma}\label{theo1}
  Under the MS-SVCJ model, the price of a European call option with strike price $K$, maturity $T$, and initial price $S_{0}$ can be written as
\vspace*{-6pt}
 \begin{equation}\label{expre3}
C=\sum_{n=0}^{+\infty}p(N_{T}=n)\sum_{\omega\in \Omega}p(\omega)\mathbb{E}_{\mathscr{J}_{n}}(e^{-rT}\mathbb{E}(S_{T}(\mathscr{J}_{n},\omega)-K)^{+})
 \end{equation}
where $N_{T}$ is the number of jumps in the asset price up to $T$; $\mathscr{J}_{n}:=(J_{1},\cdots,J_{n})$ is an $n$-dimensional random vector of $n$ jump sizes; $\Omega$ is the sample path space of $\{\sigma_{k}\}$; $S_{T}(\mathscr{J}_{n},\omega)$ is the asset price at the maturity $T$ given $\mathscr{J}_{n}$ and $\omega$.
\end{lemma}

Before deriving a tractable solution, we need the probability distribution of $S_{T}(\mathscr{J}_{n},\omega)$. Using the lemma in \cite{RN361}, we can derive the following  (see Appendix \ref{app} details):

\vspace*{-6pt}
\begin{equation}\label{equa}
  W(\mathscr{J}_{n}, \omega):=\ln\frac{S_{T}(\mathscr{J}_{n},\omega)}{S_{0}\prod\limits_{i=1}^{n}J_{i}}\Bigg|(\mathscr{J}_{n},\omega)\sim
  \mathcal{N}\left(\left(r-\lambda\zeta-\frac{\mathscr{V}(\mathscr{J}_{n}, \omega)}{2}\right)T, \mathscr{V}(\mathscr{J}_{n}, \omega)T \right),
\end{equation}
where $\prod\limits_{i=1}^{n}J_{i}$ is the cumulative effect of $n$ jumps in the asset price with $\prod\limits_{i=1}^{0}J_{i}=1$, $\zeta\equiv\mathbb{E}(J_{i}-1)$, and $\mathscr{V}(\mathscr{J}_{n}, \omega)$ is the realization of $\mathscr{V}$ given $\omega$ and $\mathscr{J}_{n}$,
$$
\mathscr{V}(\mathscr{J}_{n},\omega)=\frac{1}{T}\int_{0}^{T}\hat{\sigma}_{t}^{2}(\mathscr{J}_{n},\omega)dt
      =V+\frac{1}{T}\int_{0}^{T}\sum_{i=1}^{n}f(J_{i},t,t_{i})dt
      =|\omega|+\frac{1}{T}\sum_{i=1}^{n}f_{1}(J_{i})\int_{0}^{T}f_{2}(t,t_{i})dt.
$$
We now focus our attention on the quantity $\mathscr{V}(\mathscr{J}_{n},\omega)$. For simplicity and technical convenience, we assume that all jumps  during the interval $[T-\Delta,T]$ occur at the beginning of the interval at time $T-\Delta$; the impact of this jump time assumption on the option price is negligible, as discussed in Appendix \ref{relerr} and illustrated numerically in  Section \ref{subsec1}. Hence,
\begin{equation}\label{equa5}
\begin{aligned}
  \mathscr{V}(\mathscr{J}_{n},\omega)&=|\omega|+\hat{b}\sum_{i=1}^{n}\ln^{2}(J_{i}), \\
  \hat{b}&=\frac{b(1-e^{-\beta\Delta})}{T\beta}. \\
\end{aligned}
\end{equation}

From Lemma $\ref{theo1}$, combined with Equations $(\ref{equa})$ and $(\ref{equa5})$, we have:
  \begin{equation}\label{equa1}
    \begin{aligned}
      e^{-rT}\mathbb{E}(S_{T}(\mathscr{J}_{N_{T}},\omega)-K)^{+}&= e^{-rT}\mathbb{E}(S_{0}e^{W(\mathscr{J}_{N_{T}},\omega)}\cdot e^{\sum\limits_{i=1}^{N_{T}}\ln(J_{i})}-K)^{+}\\
      &= \mathbb{BS}(S_{0}e^{-\lambda\zeta T+\sum\limits_{i=1}^{N_{T}}\ln(J_{i})},|\omega|+\hat{b}\sum\limits_{i=1}^{N_{T}}\ln^{2}(J_{i}),r,T,K)\\
    \end{aligned}
  \end{equation}
where $\mathbb{BS}(S, \sigma^2, r, T, K)$ is the classical Black-Scholes formula as a function of initial stock price $S$, volatility $\sigma$, risk-free rate $r$, maturity $T$, and strike price $K$.
Substituting Equation $(\ref{equa1})$ into Equation $(\ref{expre3})$ yields the price of a European call option (see Appendix \ref{tho} for proof).
\begin{theorem}\label{the1}
Under the MS-SVCJ model, the price of a European call option with strike price $K$, maturity $T$, and initial underlying asset price $S_{0}$ is given by
\begin{equation*}\label{ss}
\begin{aligned}
C &=\sum_{n=0}^{+\infty}p(N_{T}=n)\sum_{v \in \Psi}p_{V}(v)C_{n}(v),\\
\end{aligned}
\end{equation*}
where $C_{n}(v) = \mathbb{E}_{\Xi_{n}}(\mathbb{BS}(S_{0}e^{-\lambda\zeta T+X_{n}},v+\hat{b}Y_{n},r,T,K))$, $\Xi_{n}:=(X_{n},Y_{n}):=(\sum\limits_{i=1}^{n}\ln(J_{i}),\sum\limits_{i=1}^{n}\ln^{2}(J_{i}))$.
\end{theorem}
$\Xi_{n}$ in Theorem \ref{the1} is a bivariate random variable, so the $n$-dimensional integral $\mathbb{E}_{\mathscr{J}_{n}}(\cdot)$ in Lemma $\ref{theo1}$ has been replaced by a double integral $\mathbb{E}_{\Xi_{n}}(\cdot)$ in Theorem \ref{the1}. In many special cases, e.g., when the jump distribution is lognormal, the probability distribution of $\Xi_{n}$ can be expressed explicitly, so that $C_{n}(Z)$ can be easily computed (see Appendix \ref{apdf} for proof).
\begin{proposition}\label{pro3}
   For $\ln(J_{i})\stackrel{\text{i.i.d.}}{\sim}\mathcal{N}\left(\mu,\varepsilon^{2}\right)$, $\zeta=\mathbb{E}(J_{i}-1)=e^{\mu+\frac{\varepsilon^{2}}{2}}-1$, the probability density function of $(X_{n},Y_{n}):=(\sum\limits_{i=1}^{n}\ln(J_{i}),\sum\limits_{i=1}^{n}\ln^{2}(J_{i}))$ is given by
  \begin{equation*}\label{expg}
    g\left(x,y\right)=
    \begin{cases}
      \frac{1}{\sqrt{2\pi}\sqrt{n\varepsilon^{2}}} e^{-\frac{\left(x-n\mu\right)^{2}}{2n\varepsilon^{2}}}\frac{1}{\Gamma\left(\frac{n-1}{2}\right)2^{\frac{n-1}{2}}}\left(\frac{y-\frac{x^2}{n}}{\varepsilon^2}\right)^{\frac{n-1}{2}-1}
      e^{-\frac{y-\frac{x^2}{n}}{2\varepsilon^2}} /\varepsilon^2, & y\geq\frac{x^2}{n}, n\geq2\\
      \frac{1}{\sqrt{2\pi\varepsilon^{2}}} e^{-\frac{\left(x-\mu\right)^{2}}{2\varepsilon^{2}}}, & y=x^{2}, n=1\\
      0, & \mbox{otherwise.}
    \end{cases}
  \end{equation*}
\end{proposition}
For lognormally distributed jump sizes, Proposition \ref{pro3} provides an analytical expression for the option price, different from the traditional solution using numerical Fourier transform inversion. \par

Moreover, under a general jump size distribution, the following corollary gives the option price for the MS-SVJ model, a special case of MS-SVCJ model (see Appendix \ref{acor1} for proof).

\begin{corollary}\label{cor1}
  Under the Markov switching stochastic volatility jump-diffusion (MS-SVJ) model, which can be recovered by setting $f(\cdot,\cdot,\cdot)=0$ in the MS-SVCJ model, and the jump size $J_{t}$ follows a general distribution, the price of a European call option is given by
$
    C=\sum\limits_{v \in \Psi}p_{V}(v)C_{jd}(v),
$
where $C_{jd}(\cdot)$ is the European call option price under the jump-diffusion model.
\end{corollary}

Thus, given the probability distribution $\{p_{V}(\cdot)\}$, the option price only depends on $C_{jd}(\cdot)$. When $\ln(J_{t})$ follows a normal distribution, a mixed-exponential distribution, or a general discrete distribution, $C_{jd}(\cdot)$ can be obtained via various approaches, such as \cite{RN342}, \cite{RN1250}, \cite{RN1249}, \cite{RN249}, or \cite{RN247}. Hence, under the MS-SVJ model with the above jump size distribution, we also provide an analytical solution for the option price.\par
Moreover, we can provide an analytical option price under the MS-SV model, which overcomes the drawbacks of \cite{RN343}, \cite{RN250}, \cite{RN199}.
\begin{corollary}\label{cor2}
  Under the Markov switching stochastic volatility (MS-SV) model, which can be recovered by setting $f(\cdot,\cdot,\cdot)=0$ and the Poisson intensity $\lambda=0$ in the MS-SVCJ model, the price of a European call option is given by
$
    C=\sum\limits_{v\in \Psi}p_{V}(v)\mathbb{BS}(S_{0},v,r,T,K).
$
\end{corollary}
{\it Proof.}
 When $f(\cdot,\cdot,\cdot)=0$ and the Poisson intensity $\lambda=0$, we have $\hat{b}=0$, $p(N_{T}=0)=1$, $p(N_{T}=n)=0, n\geq 1$, and $C_{0}(Z) =\mathbb{BS}(S_{0},Z,r,T,K)$.
\Halmos
\endproof

This result is similar to \cite{RN361}, who provided a formal solution but did not specify the distribution of AIV needed for computing the option price.\par

\subsection{American-Style Option Pricing}
By leveraging our analytical solution for European option, we can provide an efficient approximation to the price of an American-style option using the approach proposed by \cite{RN1247}. By converting the price of an American-style option to the price of a portfolio of European options, \cite{RN1247} designed algorithms to provide an upper bound and a lower bound for the price of an American-style option, where early-exercise opportunities were restricted to discrete points $0=t_{0}<t_{1}<\cdots<t_{N}=T$.

In order to apply the algorithms, for every interval $[t_{i},t_{i+1}]$, one only needs to compute two critical variables: $V_{t_{i}}(S_{i},K_{i},t_{i+1}-t_{i})$ and $\frac{\partial}{\partial S_{i}} V_{t_{i}}(S_{i},K_{i},t_{i+1}-t_{i})$, where $V_{t_{i}}(S_{i},K_{i},t_{i+1}-t_{i})$ is the European call option value with initial asset price $S_{i}$, strike price $K_{i}$ and maturity $t_{i+1}-t_{i}$. For the asset price following geometric Brownian motion and the Merton jump-diffusion model, \cite{RN1247} provided a tractable expression for $V_{t_{i}}(S_{i},K_{i},t_{i+1}-t_{i})$. \par

When the volatility $\sigma_{t}$ follows the MS process, at any given time $t_{i}$, volatility is a random variable with value space $\{u_{1},u_{2},\cdots,u_{m}\}$ and corresponding probability $\pi_{t_{i}}(.)$. Hence we have,
\begin{equation}\label{equ322}
\begin{aligned}
V_{t_{i}}(S_{i},K_{i},t_{i+1}-t_{i}) & = \sum_{\sigma \in \{u_1,\cdots,u_m\}}\pi_{t_{i}}(\sigma)C(S_{i},K_{i},t_{i+1}-t_{i},\sigma), \\
\frac{\partial}{\partial S_{i}}V_{t_{i}}(S_{i},K_{i},t_{i+1}-t_{i}) &=\sum_{\sigma \in \{u_1,\cdots,u_m\}}\pi_{t_{i}}(\sigma)\frac{\partial}{\partial S_{i}} C(S_{i},K_{i},t_{i+1}-t_{i},\sigma),
\end{aligned}
\end{equation}
 where $C(S_{i},K_{i},t_{i+1}-t_{i},\sigma)$ is the European call option with initial asset price $S_{i}$, strike price $K_{i}$, maturity $t_{i+1}-t_{i}$ and initial status $\sigma$ for the MS process in $[t_{i},t_{i+1}]$. \par

 Theorem \ref{the1}, Corollary \ref{cor1}, and Corollary \ref{cor2} provide detailed solution for $C(S_{i},K_{i},t_{i+1}-t_{i},\sigma)$ in $(\ref{equ322})$ for MS-SVCJ, MS-SVJ, and MS-SV models, respectively. As an example, the detailed expressions under the MS-SVCJ model can be written as follows by substituting Theorem \ref{the1} into (\ref{equ322}):
 \begin{equation}\label{equ321}
\begin{aligned}
V_{t_{i}}(S_{i},K_{i},t_{i+1}-t_{i}) & = \sum_{\sigma \in \{u_1,\cdots,u_m\}}\pi_{t_{i}}(\sigma)\sum_{n=0}^{+\infty}p(N_{t_{i+1}-t_{i}}=n)\sum_{v \in \Psi(\sigma)}p_{V}(v)C_{n}(S_{i},K_{i},t_{i+1}-t_{i},v), \\
\frac{\partial}{\partial S_{i}}V_{t_{i}}(S_{i},K_{i},t_{i+1}-t_{i}) &=\sum_{\sigma \in \{u_1,\cdots,u_m\}}\pi_{t_{i}}(\sigma) \sum_{n=0}^{+\infty}p(N_{t_{i+1}-t_{i}}=n)\sum_{v \in \Psi(\sigma)}p_{V}(v)\frac{\partial}{\partial S_{i}}C_{n}(S_{i},K_{i},t_{i+1}-t_{i},v),
\end{aligned}
\end{equation}
where $\Psi(\sigma)$ highlights that AIV is dependent on the initial status $\sigma$ for the MS process, and
$C_{n}(S,K,T,v)=\mathbb{E}_{\Xi_{n}}(\mathbb{BS}(Se^{-\lambda\zeta T+X_{n}},v+\hat{b}Y_{n},r,T,K))$. \par

For the MS-SV model, we have
 \begin{equation}\label{equ32}
\begin{aligned}
V_{t_{i}}(S_{i},K_{i},t_{i+1}-t_{i}) & = \sum_{\sigma \in \{u_1,\cdots,u_m\}}\sum_{v\in \Psi(\sigma)}\pi_{t_{i}}(\sigma)p_{V}(v)\mathbb{BS}(S_{i},v,r,t_{i+1}-t_{i},K_{i}), \\
\frac{\partial}{\partial S_{i}}V_{t_{i}}(S_{i},K_{i},t_{i+1}-t_{i}) &=\sum_{\sigma \in \{u_1,\cdots,u_m\}}\sum_{v\in \Psi(\sigma)}\pi_{t_{i}}(\sigma)p_{V}(v)\frac{\partial}{\partial S_{i}} \mathbb{BS}(S_{i},v,r,t_{i+1}-t_{i},K_{i}).
\end{aligned}
\end{equation}

Up to now, we have provided an analytical solution for option price under the proposed model with the lognormal jump size distribution. Practical application requires calculating the probability distribution of $V$ efficiently, which is addressed in the next section.

\section{An Efficient Algorithm for MS-process AIV}
\vspace*{-6pt}
After observing that complete enumeration (CE) based on the definition of $V$ is intractable due to the enormous computation time and consumed memory, we develop an efficient algorithm for the probability distribution $\{p_{V}(\cdot)\}$ called the recursive recombination (RR) algorithm.

\vspace*{-6pt}
\vspace*{-6pt}
\subsection{Complete Enumeration (CE)}
Based on the definition of $V$ in Section \ref{subsection2}, a complete enumeration (CE) algorithm for $\Psi$ and $\{p_{V}(\cdot)\}$ would traverse all sample paths $\omega$ and generate $(|\omega|,p(\omega))$; then
collect distinct values $|\omega|$ to get $\Psi$ and sum probabilities $p(\omega)$ with $|\omega|=v \in \Psi$ to get $p_{V}(v)$.
However, for the MS process with $m$ states, the complexity of CE is clearly $O(m^{L})$, exponential in the number of time steps. Hence, to efficiently derive $\Psi$ and $\{p_{V}(\cdot)\}$, we propose the RR algorithm.\par

\vspace*{-6pt}
\subsection{Recursive Recombination (RR) Algorithm}
\vspace*{-6pt}
Recall from Section $\ref{subsection2}$, for the MS process $\{\sigma_{k}\}$, the initial state $\sigma_{0}$ is fixed. Here, we define a subsample path
$\omega_{l}=\left[\sigma_{0},\sigma_{1},\sigma_{2},\ldots,\sigma_{l-2},\sigma_{l-1},\sigma_{l}\right]$ of $\{\sigma_{k}\}$ up to step $l$, as the first $l+1$ elements of a sample path $\omega$, with weight $|\omega_{l}|$ and corresponding probability $p(\omega_{l})$:
\begin{equation*}
\begin{aligned}
|\omega_{l}| & =\frac{\sigma_{0}^{2}+\sigma_{1}^{2}+\sigma_{2}^{2}+\ldots+\sigma_{l-2}^{2}+\sigma_{l-1}^{2}}{l}, \\
p(\omega_{l}) & =p_{\sigma_{0}\sigma_{1}}p_{\sigma_{1}\sigma_{2}}....p_{\sigma_{l-2}\sigma_{l-1}}p_{\sigma_{l-1}\sigma_{l}}.
\end{aligned}
\end{equation*}
In addition, we also denote the set of all subsample paths $\omega_{l}$ as $\Omega_{l}$, which is the subsample path space for $\{{\sigma_{k}}\}$ up to step $l$. \par
For a subsample path $\omega_{l}$, we extract three fundamental features: the weight $|\omega_{l}|$, the length $l$ and the last element $\sigma_{l}$, which generate a triple $[|\omega_{l}|,l,\sigma_{l}]$. Thus, $[|\omega_{l}|,l,\sigma_{l}]$ is a random variable with value space
$
   \Psi_{l}=\{[x,l,\sigma_{l}]: x=|\omega_{l}|, \sigma_{l} \text{ is the last element of }  \omega_{l}, \omega_{l} \in \Omega_{l} \},
$
and corresponding probability
$
p([x,l,\sigma_{l}])=
  \sum\limits_{\omega_{l}\in \Omega_{l}:|\omega_{l}|=x, \omega_{l} \text{ ends with } \sigma_{l}}p(\omega_{l}), [x,l,\sigma_{l}] \in \Psi_{l}$.

Now we can relate $[|\omega_{L}|,L,\sigma_{L}]$ to the random variable $V$:
  \begin{equation}\label{equa11}
    \begin{aligned}
       \Psi &=  \{v: [v, L, \sigma_{L}] \in \Psi_{L}\}, \\
       p_{V}(v) & =\sum_{[v,L,\sigma_{L}]\in \Psi_{L}}p([v, L, \sigma_{L}]). \\
     \end{aligned}
  \end{equation}

\begin{figure}[tb]
  \centering
\begin{tikzpicture}[scale=0.8]
\draw[thick] (0,0) -- (18,0);
\draw[thick] (0.5,0) -- (0.5,-0.15) node [below]{0};
\draw[thick] (5.5,0) -- (5.5,-0.15) node [below]{1};
\draw[thick] (10.5,0) -- (10.5,-0.15) node [below]{2};
\draw[thick] (15.5,0) -- (15.5,-0.15) node [below]{3};

\node (a1) at (0.5,4.5) {$[0.4]$};
\node (b1) at (5.5,6.5) {$[0.4,0.4]$};
\node (b2) at (5.5,2.5) {$[0.4,0.2]$};
\node (c1) at (10.5, 7.5) {$[0.4,0.4,0.4]$};
\node (c2) at (10.5, 5.5) {$[0.4,0.4,0.2]$};
\node (c3) at (10.5, 3.5) {$[0.4,0.2,0.4]$};
\node (c4) at (10.5, 1.5) {$[0.4,0.2,0.2]$};
\node (d1) at (15.5, 8) {$[0.4,0.4,0.4,0.4]$};
\node (d2) at (15.5, 7) {$[0.4,0.4,0.4,0.2]$};
\node (d3) at (15.5, 6) {$[0.4,0.4,0.2,0.4]$};
\node (d4) at (15.5, 5) {$[0.4,0.4,0.2,0.2]$};
\node (d5) at (15.5, 4) {$[0.4,0.2,0.4,0.4]$};
\node (d6) at (15.5, 3) {$[0.4,0.2,0.4,0.2]$};
\node (d7) at (15.5, 2) {$[0.4,0.2,0.2,0.4]$};
\node (d8) at (15.5, 1) {$[0.4,0.2,0.2,0.2]$};

\draw[thick] (a1) -- (b1);
\draw[thick] (a1) -- (b2);
\draw[thick] (b1) -- (c1);
\draw[thick] (b1) -- (c2);
\draw[thick] (b2) -- (c3);
\draw[thick] (b2) -- (c4);
\draw[thick] (c1) -- (d1);
\draw[thick] (c1) -- (d2);
\draw[thick] (c2) -- (d3);
\draw[thick] (c2) -- (d4);
\draw[thick] (c3) -- (d5);
\draw[thick] (c3) -- (d6);
\draw[thick] (c4) -- (d7);
\draw[thick] (c4) -- (d8);
\end{tikzpicture}
\vspace*{-6pt}
\caption{$\Omega_{l}$, Subsample Path Space of $\omega_{l}$ Up to Step $L=3$}\label{figure3}
\end{figure}
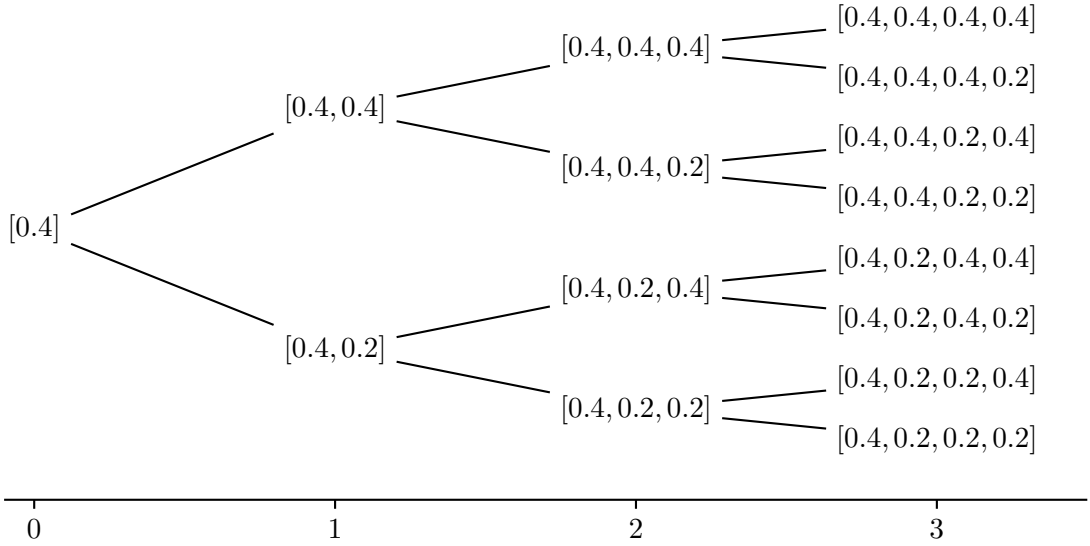

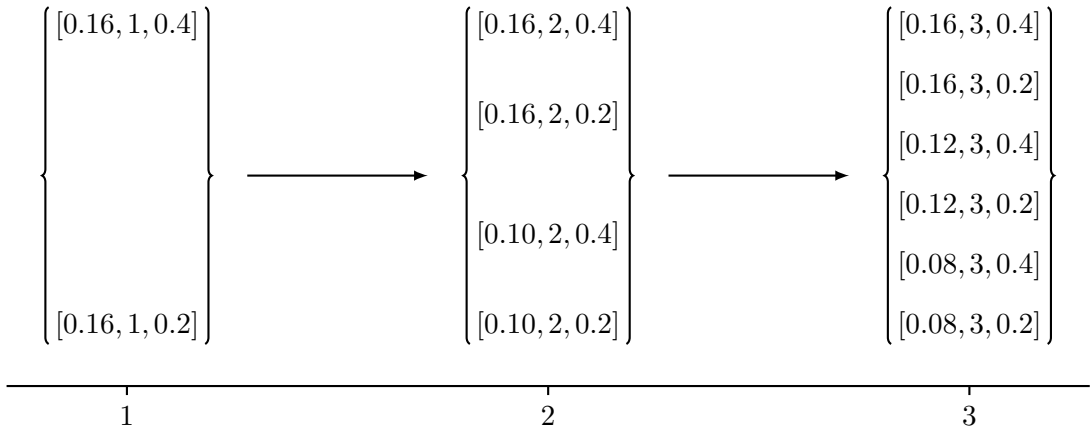
\begin{figure}[tb]
  \centering
\begin{tikzpicture}[scale=0.8]
\draw[thick] (0,0) -- (18,0);
\draw[thick] (2,0) -- (2,-0.15) node [below]{1};
\draw[thick] (9,0) -- (9,-0.15) node [below]{2};
\draw[thick] (16,0) -- (16,-0.15) node [below]{3};

\node (a1) at (2,6) {$[0.16,1,0.4]$};
\node (a2) at (2,1) {$[0.16,1,0.2]$};
\node (b1) at (9, 6) {$[0.16,2,0.4]$};
\node (b2) at (9, 4.5) {$[0.16,2,0.2]$};
\node (b3) at (9, 2.5) {$[0.10,2,0.4]$};
\node (b4) at (9, 1) {$[0.10,2,0.2]$};
\node (c1) at (16, 6) {$[0.16,3,0.4]$};
\node (c2) at (16, 5) {$[0.16,3,0.2]$};
\node (c3) at (16, 4) {$[0.12,3,0.4]$};
\node (c4) at (16, 3) {$[0.12,3,0.2]$};
\node (c5) at (16, 2) {$[0.08,3,0.4]$};
\node (c6) at (16, 1) {$[0.08,3,0.2]$};

\draw[decorate,decoration={brace},thick] (0.7,0.7) -- (0.7,6.3);
\draw[decorate,decoration={brace,mirror},thick] (3.3,0.7) -- (3.3,6.3);
\draw[decorate,decoration={brace},thick] (7.7,0.7) -- (7.7,6.3);
\draw[decorate,decoration={brace,mirror},thick] (10.3,0.7) -- (10.3,6.3);
\draw[decorate,decoration={brace},thick] (14.7,0.7) -- (14.7,6.3);
\draw[decorate,decoration={brace,mirror},thick] (17.3,0.7) -- (17.3,6.3);

\draw[thick,-latex] (4,3.5) -- (7,3.5);
\draw[thick,-latex] (11,3.5) -- (14,3.5);
\end{tikzpicture}
\vspace*{-6pt}
\caption{$\Psi_{l}$, Value Space of $[|\omega_{l}|,l,\sigma_{l}]$ Up to Step $L=3$}\label{figure4}
\end{figure}

We provide an example with state set $\{0.2,0.4\}$, number of time steps $L=3$, and initial state $\sigma_{0}=0.4$. In Figure \ref{figure3}, all the subsample paths with the same $l$ constitute $\Omega_{l}$. Taking $l=3$ as an example, eight subsample paths generate $\Omega_{3}=\{[0.4,0.4,0.4,0.4],\cdots,[0.4,0.2,0.2,0.2]\}$. Based on $\Omega_{l}$ in Figure \ref{figure3}, Figure \ref{figure4} presents the corresponding value space $\Psi_{l}$. As an example with $l=3$, the different subsample paths $[0.4,0.4,0.2,0.4]$ and $[0.4,0.2,0.4,0.4]$ are distinct elements in $\Omega_{3}$, but since they have the same weight $\frac{0.4^2+0.4^2+0.2^2}{3}=\frac{0.4^2+0.2^2+0.4^2}{3}=0.12$, the same length and last element, then they correspond to the same element $[0.12,3,0.4]$ in $\Psi_{3}$. Finally, based on Equation $(\ref{equa11})$, we have $\Psi=\{0.16,0.12,0.08\}$, which is consistent with the definition of $V$.\par

Now we present a recursive algorithm to obtain $\Psi_{L}$ and $p([x,L,\sigma_{L}])$. Since the initial state $\sigma_{0}$ is fixed, for $l=1$, we have $\Psi_{1}= \{[\sigma_{0}^{2},1,\sigma_{1}]:  \sigma_{1} \in \{u_{1},\cdots,u_{m}\}\}$ and $p([\sigma_{0}^{2},1,\sigma_{1}])= p_{\sigma_{0}\sigma_{1}}$. \par

The main recursive step is based on the following proposition, where $\omega_{l+1}$ can be generated from $\omega_{l}$ by taking a step forward.
\begin{proposition}\label{theo4}
The value space and the probability of $[|\omega_{l}|,l,\sigma_{l}]$ follow the recursive relationship:
\begin{equation*}
  \begin{aligned}
    \Psi_{l+1} &=\{[z,l+1,\sigma_{l+1}]: z=\frac{x\cdot l+\sigma_{l}^{2}}{l+1}, [x,l,\sigma_{l}]\in \Psi_{l}, \sigma_{l+1} \in \{u_{1},\cdots,u_{m}\}\}, \\
    p([z,l+1,\sigma_{l+1}]) &=\sum_{[x,l,\sigma_{l}]\in \Psi_{l}: x=(z\cdot (l+1)-\sigma_{l}^{2})/{l}}p([x,l,\sigma_{l}])p_{\sigma_{l}\sigma_{l+1}}.
  \end{aligned}
\end{equation*}
\end{proposition}\par

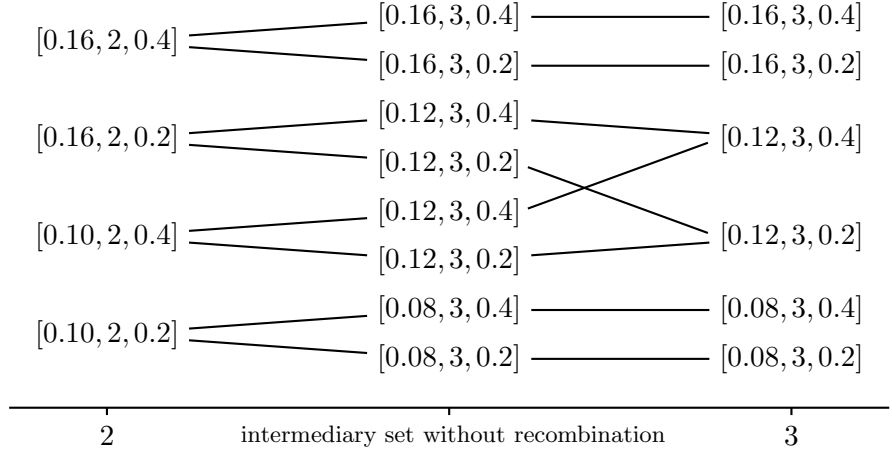
\begin{figure}[tb]
  \centering
\begin{tikzpicture}[scale=0.65]
\draw[thick] (0,0) -- (18,0);
\draw[thick] (2,0) -- (2,-0.15) node [below]{2};
\draw[thick] (9,0) -- (9,-0.15) node [below]{\mbox{\footnotesize intermediary set without recombination}};
\draw[thick] (16,0) -- (16,-0.15) node [below]{3};

\node (a1) at (2,7.5) {$[0.16,2,0.4]$};
\node (a2) at (2,5.5) {$[0.16,2,0.2]$};
\node (a3) at (2,3.5) {$[0.10,2,0.4]$};
\node (a4) at (2,1.5) {$[0.10,2,0.2]$};
\node (b1) at (9, 8) {$[0.16,3,0.4]$};
\node (b2) at (9, 7) {$[0.16,3,0.2]$};
\node (b3) at (9, 6) {$[0.12,3,0.4]$};
\node (b4) at (9, 5) {$[0.12,3,0.2]$};
\node (b41) at (10.4, 5) {};
\node (b5) at (9, 4) {$[0.12,3,0.4]$};
\node (b51) at (10.4, 4) {};
\node (b6) at (9, 3) {$[0.12,3,0.2]$};
\node (b7) at (9, 2) {$[0.08,3,0.4]$};
\node (b8) at (9, 1) {$[0.08,3,0.2]$};
\node (c1) at (16, 8) {$[0.16,3,0.4]$};
\node (c2) at (16, 7) {$[0.16,3,0.2]$};
\node (c3) at (16, 5.5) {$[0.12,3,0.4]$};
\node (c31) at (14.5, 5.5) {};
\node (c4) at (16, 3.5) {$[0.12,3,0.2]$};
\node (c41) at (14.5, 3.5) {};
\node (c5) at (16, 2) {$[0.08,3,0.4]$};
\node (c6) at (16, 1) {$[0.08,3,0.2]$};

\draw[thick] (a1) -- (b1);
\draw[thick] (a1) -- (b2);
\draw[thick] (a2) -- (b3);
\draw[thick] (a2) -- (b4);
\draw[thick] (a3) -- (b5);
\draw[thick] (a3) -- (b6);
\draw[thick] (a4) -- (b7);
\draw[thick] (a4) -- (b8);
\draw[thick] (b1) -- (c1);
\draw[thick] (b2) -- (c2);
\draw[thick] (b3) -- (c3);
\draw[thick] (b41) -- (c41);
\draw[thick] (b51) -- (c31);
\draw[thick] (b6) -- (c4);
\draw[thick] (b7) -- (c5);
\draw[thick] (b8) -- (c6);
\end{tikzpicture}
\vspace*{-8pt}
\caption{Recursion Recombination Relationship From $\Psi_{2}$ to $\Psi_{3}$}\label{figure5}
\end{figure}

Continuing with the example above, Figure \ref{figure5} illustrates recursion and recombination from from $\Psi_{2}$ to $\Psi_{3}$. We start from $\Psi_{2}$, then take a step forward to generate an intermediary set without recombination. For example, $[0.16,2,0.2]\in\Psi_{2}$ and $[0.10,2,0.4]\in\Psi_{2}$ taking a step forward generate $\{[0.12,3,0.4],[0.12,3,0.2]\}$ and $\{[0.12,3,0.4],[0.12,3,0.2]\}$, respectively. After recombining the same element in the intermediary set, for example $[0.12,3,0.4]$ and $[0.12,3,0.2]$, we obtain $\Psi_{3}$.

Table $\ref{table1}$ provides the RR algorithm for the value space $\Psi$ and the probability distribution $\{p_{V}(\cdot)\}$. In terms of computational complexity, we have the following result (see Appendix \ref{comrsp} for proof):
\begin{proposition}\label{prop5}
  The total number of distinct triples $[x,l,\sigma_{l}]$ from step $1$ to step $L$ is bounded by $m\binom{L-1+m}{m}$; hence the complexity of the RR algorithm is $O\left(L^{m}\right)$.
\end{proposition}
Since $L\gg m$ in settings of practical interest, the RR algorithm should be far superior to CE in terms of computation, which is confirmed in the next section.

\begin{table}[tb]
  \centering
  \caption{ RR Algorithm: Obtaining the Value Space $\Psi$ and the Probability Distribution $\{p_{V}(\cdot)\}$}
  \label{table1}
  \begin{tabular}{l}
  \toprule[2pt]
  \textbf{Input:} \tabincell{l}{State set $\{u_{1},\cdots,u_{m}\}$, transition probabilities $p_{ij}$, $i,j\in\{1,\cdots,m\},$\\[-8pt]
  initial state $\sigma_{0}$, number of time steps $L$.} \\ \midrule[1pt]
  \textbf{Initialization:} \tabincell{l}{Set $\Psi_{1}=\{[\sigma_{0}^{2},1,\sigma_{1}]: \sigma_{1} \in \{u_{1},\cdots,u_{m}\}\}$,\vspace{0.2cm} \\[-10pt]
  $p([\sigma_{0}^{2},1,\sigma_{1}]) =p_{\sigma_{0}\sigma_{1}}$.}\\ \midrule[1pt]
  \textbf{Recursion:} \tabincell{l}{ For $l=1$ to $L-1$\\[-8pt]
  ~~~~Let $\Psi_{l+1}=\{[z,l+1,\sigma_{l+1}]: z=\frac{x\cdot l+\sigma_{l}^{2}}{l+1}, [x,l,\sigma_{l}]\in \Psi_{l}, \sigma_{l+1} \in \{u_{1},\cdots,u_{m}\}\}$,\vspace{0.2cm}\\[-10pt]
  ~~~~$p([z,l+1,\sigma_{l+1}])=\sum\limits_{[x,l,\sigma_{l}]\in \Psi_{l}: x=(z\cdot (l+1)-\sigma_{l}^{2})/{l}}p([x,l,\sigma_{l}])p_{\sigma_{l}\sigma_{l+1}}$.} \\ \midrule[1pt]
  \textbf{Output:}  ~~~~\tabincell{l}{$\Psi=\{x: [x, L, \sigma_{L}] \in \Psi_{L}\}$,\vspace{0.2cm} \\[-10pt]
  $p_{V}(v)=\sum\limits_{[x,L,\sigma_{L}]\in \Psi_{L}:x=v}p(x, L, \sigma_{L})$. }\\
  \bottomrule[2pt]
\end{tabular}
\end{table}\par

\vspace*{-6pt}
\section{Numerical Experiments}\label{sec1}
\vspace*{-6pt}
In this section, we first price the European call option under the proposed model. We conduct Monte Carlo simulation to assess the impact of the jump time assumption on the option price. After that, we discuss the result for the Bermudan call option pricing to show the effectiveness of our approach. Then we compare the proposed RR algorithm with CE, and the results highlight the computational superiority of the RR algorithm. Lastly, we provide an application example of fitting to real-market data. All numerical results are obtained using MATLAB (see  Appendix \ref{rrsp} for MATLAB code) on a 2.40 GHz Intel Xeon E5-2680, 128 GB RAM computer.

\vspace*{-6pt}
\subsection{European Option}\label{subsec1}
\vspace*{-6pt}
Before applying our pricing method under the proposed model, we conduct a Monte Carlo simulation to assess the impact of the assumption on jump time on the option price. The parameters for the proposed model are presented in Table \ref{table6}. Specifically, we follow \cite{RN373} to assign the values of duration $\Delta$, attenuating factor $\beta$, proportional coefficient $b$ (see  Appendix \ref{para}).

\begin{table}[tb]
  \centering
  \caption{Parameter Values for MS-SVCJ Model}
  \scalebox{0.8}[0.75]{
  \begin{threeparttable}
    \begin{tabular}{llllllll}
    \hline
    Parameter & Value &       & Parameter & Value &   & Parameter  &  Value \\
    \hline
    Maturity &   $T=0.25$    &       & Jump variance &  $\varepsilon^{2}=0.005$  &   & Initial state & $\sigma_{0}^{2}=0.04$ \\
    Strike price &   $K=55$   &      &  Max \# jumps\tnote{$\dagger$}  &  $N_{max}=10$ &   & State space  & $\sigma_{k}^{2} \in \{0.02,0.04,0.06,0.08\} $ \\
    Risk-free rate &  $r=0.05$     &       &  Time step & $\tau=0.25/30$  &   & \multirow{4}{*}{\tabincell{l}{Transition\\ probability\\ matrix}}  &  \multirow{4}{*}{  $
P=\left[
  \begin{array}{cccc}
     0.70   & 0.15   & 0.10  & 0.05 \\
     0.03  & 0.90   & 0.06  & 0.01 \\
     0.05  & 0.05  & 0.85  & 0.05 \\
     0.03  & 0.07  & 0.10  & 0.80 \\
  \end{array}
  \right]
  $}  \\
    Asset price &   $S_{0}=50$         &   & Duration & $\Delta=0.02$  &   &   &   \\
    Jump intensity &     $\lambda=3$      &   & Attenuating factor & $\beta=250$   &   &   &   \\
    Jump mean &    $\mu=-0.025$       &   & Proportional coefficient & $b=2$   &   &   &   \\
    \bottomrule
    \end{tabular}%
     \begin{tablenotes}
     \footnotesize
    \item[$\dagger$]\hspace*{-8pt} max \# jumps truncated at $N_{max}$ such that $P(N>N_{max}) < \epsilon$;
    for $\epsilon=5.5\times10^{-5}$ with $\lambda$ and $T$ values, $N_{max}=10$.
  \end{tablenotes}
 \end{threeparttable}
  \label{table6} }
\end{table}

Using an Euler approximation with $N$ equal subintervals, Monte Carlo simulation generates the asset price at maturity. For the case in Table \ref{table6}, the option prices using Theorem \ref{the1}, denoted by MS-SVCJ, and Monte Carlo simulation, denoted by MC, are summarized in Table $\ref{table7}$. For the MC column, option price, standard error, and computation time are based on 10 sets of $100,000$ paths. MC simulation is very time-consuming, with an example of $N=1500$ taking around $11$ hours.

\begin{table}[tb]
  \centering
  \caption{Option Valuation Comparison}
  \scalebox{1.00}[1.00]{
    \begin{tabular}{l|c|ccccc}
    \hline
       & MS-SVCJ & \multicolumn{4}{c}{MC Simulation} \\[-9pt]
      & & $N$=600 &750 &900  &1200 &1500 \\
    \hline
    Option Price &   0.9696 &  0.9680  & 0.9683 &  0.9684  &  0.9687   &   0.9689 \\[-9pt]
    (Std Err) &   & (.0063) & (.0044) &  (.0050)  &  (.0066)   &   (.0054) \\
    \hline
    Computation Time (seconds) &  30  & 16575 & 20213 & 25014 & 28411 & 41034\\
    \hline
    \end{tabular}
  \label{table7}
  }
\end{table}\par
As mentioned in Section \ref{subsec2} and  Appendix \ref{relerr}, the assumption on jump time will increase the volatility slightly, hence increase the option price slightly. On the other hand, the MC values monotonically increase with $N$, so the option price without the assumption lies in the interval $[0.9689,0.9696]$, bounding the relative error at less than $0.07\%=\frac{|0.9696-0.9689|}{0.9689}$ and indicating that the impact of the assumption on the option price is negligible. Moreover, the option price with the assumption stays within the $95\%$ confidence interval of all the MC values, providing further support for the reasonableness of the assumption.\par

\subsection{American-Style Option}\label{section52}
For an American-style option, we apply the secant and tangent algorithms of \cite{RN1247} to establish bounds for its price. Under the MS-SV model, the value $V_{t_{i}}(S_{i},K_{i},t_{i+1}-t_{i})$ is a weighted sum of $\mathbb{BS}(S_{i},v,r,t_{i+1}-t_{i},K_{i})$ over the possible AIVs. We use $n$ interpolation points for the asset price, where the approximation accuracy can be improved by increasing the number of interpolation points. As a comparison, we also implement the least squares Monte Carlo simulation approach of \cite{RN3931} for American option pricing, denoted by LSM, based on 10 sets of independent runs with 100,000 paths for each run. \par

We illustrate our approach with the following example from \cite{RN1247}: a three-year Bermudan call option with strike price $K=100$, exercisable every $0.5$ years. For the dynamics under the MS-SV model, we set risk-free rate $r=0.05$, dividend rate $\delta=0.04$, initial state $\sigma_{0}^{2}=0.04$, state space $\sigma_{t}^{2}\in\{0.02,0.04,0.06,0.08\}$, time step $\tau=0.5/30$ with the transition matrix $P$ shown in Table \ref{table6}. We calculate $V_{t_{i}}(S_{i},K_{i},t_{i+1}-t_{i})$ and $\frac{\partial}{\partial S_{i}} V_{t_{i}}(S_{i},K_{i},t_{i+1}-t_{i})$ using Equation $(\ref{equ32})$.

The results in Table \ref{table77} show that the upper and lower bounds derived from the secant and tangent algorithms, respectively, converge quickly with the number of interpolating points. For example, with 100 interpolating points, the upper and lower bounds are within a penny of the true price. For $n=200$, option prices with different initial asset prices via the secant and tangent algorithms all stay with the $95\%$ confidence interval of the corresponding LSM values. In addition, the computation time for LSM is orders of magnitude higher than the time for the secant and tangent approximation approaches.

\begin{table}[tb]
  \centering
  \caption{Bermudan Call Option Pricing Under MS-SV Model}
  \scalebox{1}[1]{
  \begin{threeparttable}
    \begin{tabular}{lc|rrrrr|c}
    \hline
      &   & \multicolumn{5}{c|}{Option Price} & Computation Time \\[-5pt]
       \cline{3-7}
      Algorithm& $n$ & $S_{0}$=60 & \multicolumn{1}{c}{90} & \multicolumn{1}{c}{100}  & \multicolumn{1}{c}{110} & \multicolumn{1}{c|}{140} & (seconds)\\
    \hline
    Tangent &   50 &  1.294  & 9.846 &  14.867  &  20.848   &   43.204 & 0.41\\[-2pt]
     &   100 &  1.302  & 9.861 &  14.883  &  20.861   &   43.212 &0.98\\[-2pt]
     &   200 &  1.305  & 9.864 &  14.886  &  20.864   &   43.213 &2.97\\
    \hline
     Secant &   200 &  1.307  & 9.868 &  14.890  &  20.867   &   43.215 & 1.64\\[-2pt]
     &   100 &  1.311  & 9.875 &  14.897  &  20.873   &   43.219 & 0.55\\[-2pt]
     &   50 &  1.328  & 9.904 &  14.925  &  20.899   &   43.234 & 0.23\\
     \hline
     LSM & & 1.306 & 9.866& 14.888& 20.860 &43.210 & 415\\[-10pt]
     (Std Err) & & (.016) & (.019) & (.043) & (.020) & (.074) & \\
    \hline
    \end{tabular}%
  \end{threeparttable}
  \label{table77}%
  }
\end{table}

Next we apply our approach to price the same call option under the MS-SVCJ model, which is very similar to MS-SV model with additional jumps and co-jumps. The model parameters are the same as in Table \ref{table6}, and $V_{t_{i}}(S_{i},K_{i},t_{i+1}-t_{i})$ and $\frac{\partial}{\partial S_{i}} V_{t_{i}}(S_{i},K_{i},t_{i+1}-t_{i})$ are calculated via Equation $(\ref{equ321})$. The results for $n=20$, $50$ and $100$ are illustrated in Table \ref{table771}.

\begin{table}[tb]
  \centering
  \caption{Bermudan Call Option Pricing Under MS-SVCJ Model}
  \scalebox{1}[1]{
  \begin{threeparttable}
    \begin{tabular}{lc|rrrrr|c}
    \hline
      &   & \multicolumn{5}{c|}{Option Price} & Computation Time \\[-5pt]
       \cline{3-7}
      Algorithm& $n$ & $S_{0}$=60 & \multicolumn{1}{c}{90} & \multicolumn{1}{c}{100}  & \multicolumn{1}{c}{110} & \multicolumn{1}{c|}{140} & (seconds)\\
    \hline
    Tangent &   20 &  1.970  & 11.624 &  16.815  &  22.845   &   44.864 &68\\[-2pt]
     &   50 &  2.040  & 11.723 &  16.911  &  22.932   &   44.924 &390\\[-2pt]
        &   100 &  2.050  & 11.737 &  16.924  &  22.945   &   44.933 &1504\\
    \hline
    Secant &   100 &  2.060  & 11.752 &  16.938  &  22.957   &   44.941 &895\\[-2pt]
     &   50 &  2.080  & 11.780 &  16.965  &  22.982   &   44.958 &230\\[-2pt]
     &   20 &  2.223  & 11.977 &  17.154  &  23.157   &   45.078 &39\\
     \hline
     LSM & & 2.066& 11.773 &16.957& 22.972 & 44.903 &23934\\[-10pt]
     (Std Err) & & (.028) &(.082) & (.082) & (.079) & (.095) & \\
    \hline
    \end{tabular}%
  \end{threeparttable}
  \label{table771}%
  }
\end{table}

\vspace*{39pt}
\subsection{CE versus RR}\label{section51}
\vspace*{-6pt}
We compare the computation time for CE and the RR algorithm as a function of $L$ and $m$, with the results provided in Tables \ref{tabel4} and \ref{tabel5}, respectively. The results in Table $\ref{tabel4}$ illustrate that for CE, the computation time increases exponentially with respect to the number of time steps $L$, and the consumed memory is quickly exhausted, which limits the application of CE in practice. Comparing Table $\ref{tabel4}$ with Table $\ref{tabel5}$, the improvement using the RR algorithm is significant. For example, for $m=5$, $L=40$ or $m=6$, $L=30$, the RR algorithm finishes within about 10 seconds.

\begin{table}[tb]
  \centering
  \caption{Computation Time for CE (seconds, `*' indicates out of memory)}
  \scalebox{1}[1]{
  \begin{threeparttable}
\begin{tabular}{l|cccccccc}
    \hline
\backslashbox{$m$}{$L$~~~}   & 15    & 16    & 17    & 18    & 19    & 20    & 25  &30\\
     \hline
    2     &    0.007   &   0.01    &    0.04   &   0.04   &    0.07   &    0.13   &  4.5 &151\\
    \hline
    3     &    1.3  &   4.1  &    12   &    38   &    119   &    365  &  *&*\\
    \hline
    4     &    92   &     *  &    *   &    *   &    *   &    *   &  * &*\\
    \hline
    5     &   *    &     *  &    *   &    *   &    *   &    *   &  * &*\\
    \hline
    6     &   *    &     *  &    *   &    *   &    *   &    *   & * &*\\
    \hline
    \end{tabular}
 \end{threeparttable}
  \label{tabel4}
  }
\end{table}%

\begin{table}[tb]
  \centering
  \caption{Computation Time for RR Algorithm (seconds)}
  \scalebox{1}[1]{
\begin{tabular}{l|ccccccc}
    \hline
\backslashbox{$m$}{$L$~~~}   & 20    & 25    & 30    & 35    & 40    & 45    & 50 \\
    \hline
    2     &    0.005   &    0.006   &    0.007   &    0.008   &   0.011    &   0.013    & 0.014 \\
    \hline
    3     &    0.009   &    0.013   &    0.019   &    0.024   &   0.033  &    0.040   &  0.050 \\
    \hline
    4     &    0.05   &    0.11   &    0.21   &    0.39  &    0.66   &    1.05   &  1.58 \\
    \hline
    5     &    0.27   &    0.8  &     1.9   &    4.0   &   7.3   &   12    & 19 \\
    \hline
    6     &    1.2  &     4.2  &    10   &    22   &    38   &    60   &  87 \\
    \hline
    \end{tabular}%
  \label{tabel5}
  }
\end{table}%

\begin{figure}[tb]
  \centering
  \subfigure[CE with Two States]{
  \begin{minipage}{0.48\linewidth}
  \centering
    \includegraphics[height=6.0cm,width=8.0cm]{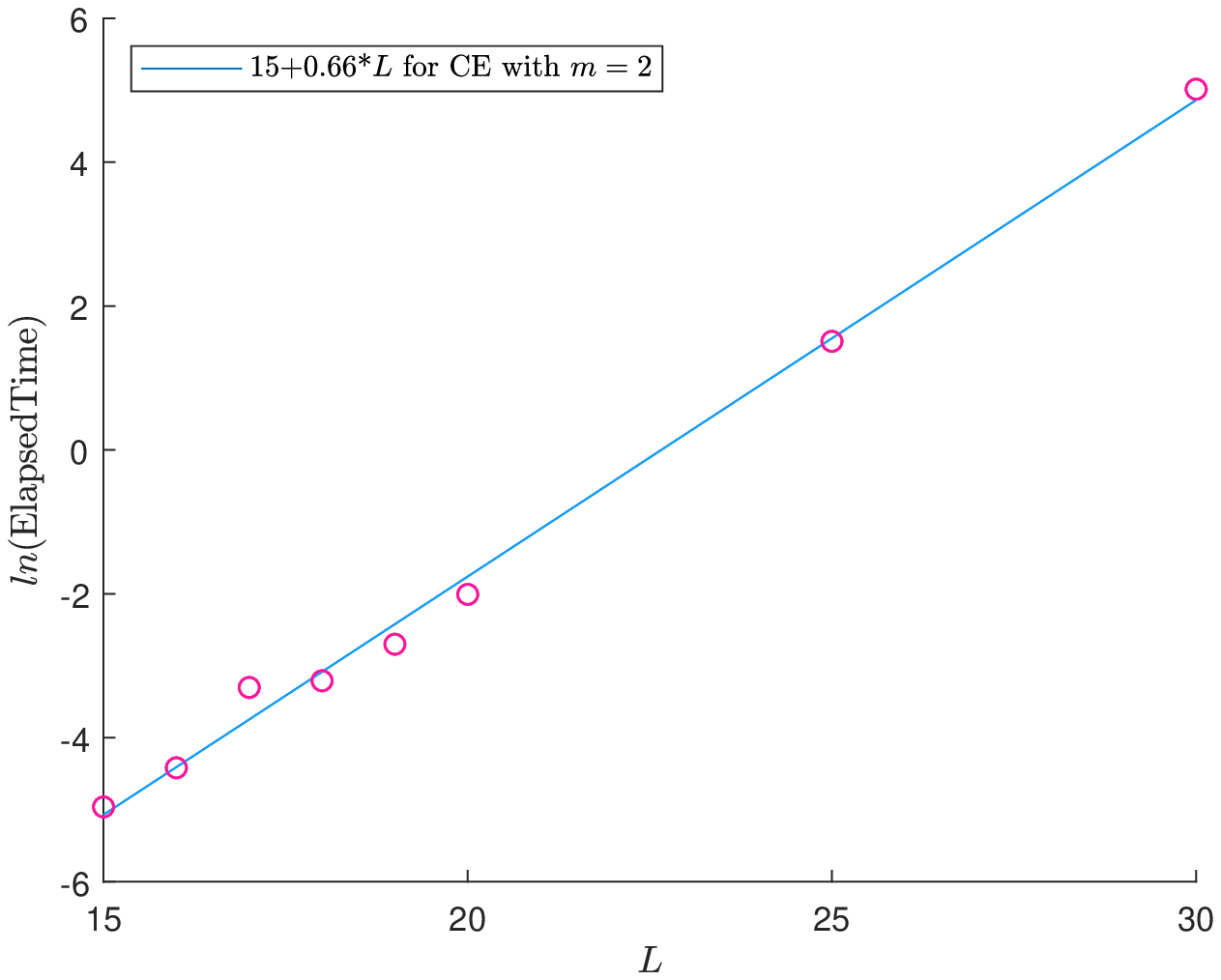}
  \end{minipage}
  }
 \subfigure[RR Algorithm with Five States]{
  \begin{minipage}{0.48\linewidth}
  \centering
    \includegraphics[height=6.0cm,width=8.0cm]{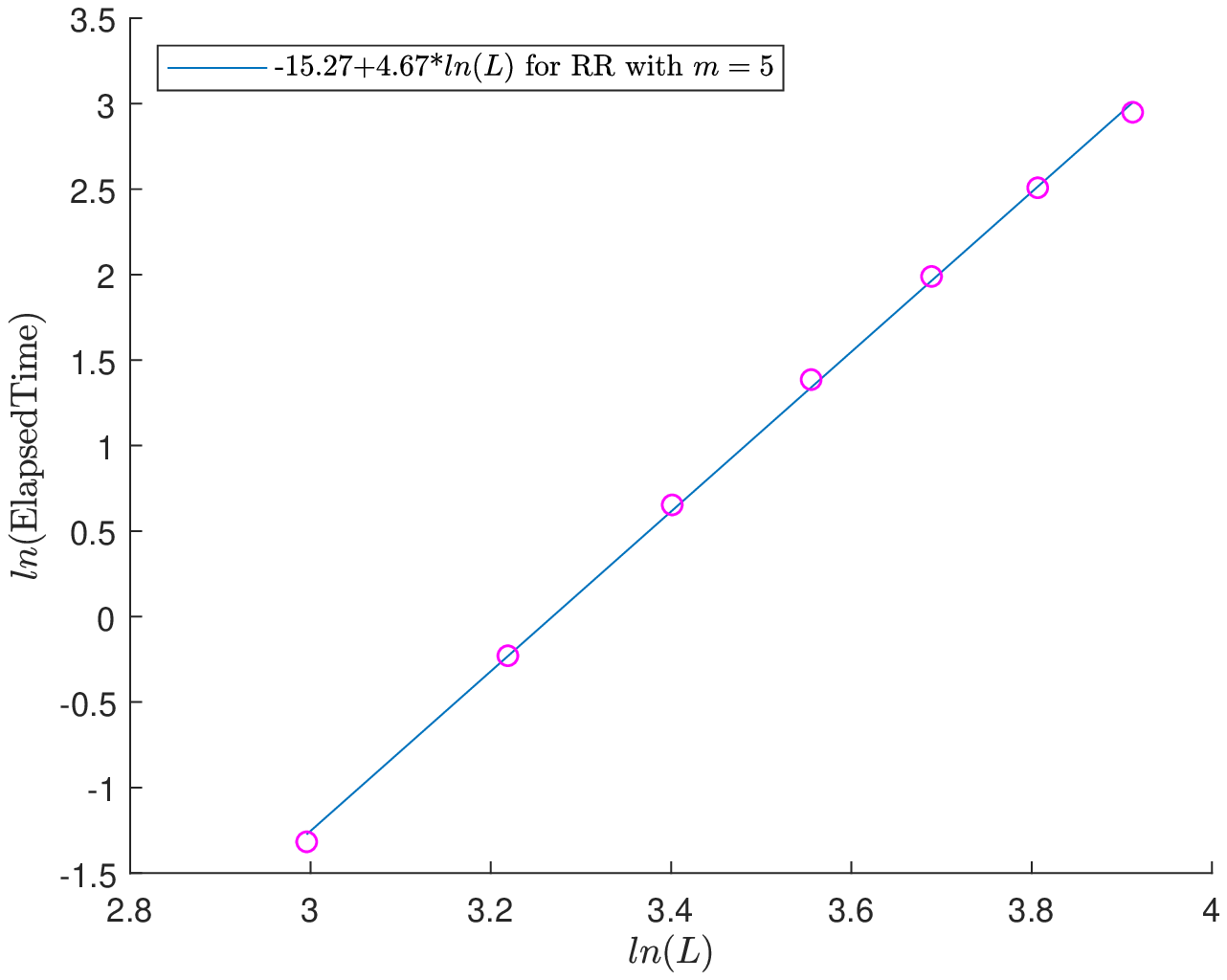}
  \end{minipage}
  }
\vspace*{-2pt}
  \caption{Computation Time as a Function of the Number of Time Steps $L$ (Log Scales)}
  \label{figure2}
\end{figure}

To illustrate the computational complexity of both algorithms, we graph the computation time as a function of the number of time steps $L$ for both algorithms. Figure \ref{figure2} shows representative plots of the performance of CE with two states and the RR algorithm with five states; additional results are provided in  Appendix \ref{plot}. The results support the theoretical computational complexity of $O(m^{L})$ for CE and $O(L^{m})$ for the RR algorithm.\par

\subsection{Application Example}\label{subs53}
\vspace*{-6pt}
We calibrate our model to market option prices after estimating the MS and PEA model parameters for the underlying asset prices using real data from Yahoo Finance consisting of the daily closing stock prices of IBM from January 1, 2005 to June 30, 2019. We estimate the jump process using the box-plot method. Specifically, we assume values of the daily log-return $r^{a}_{t}=\ln (S_{t+a}/S_{t})$ outside the range $(Q_{1}-k_{f}R_{f}, Q_{3}+k_{f}R_{f})$ constitute jumps, where $a=1/252$ is the sampling interval length for the daily price, $Q_{1}$ is the lower quartile, $Q_{3}$ is the upper quartile, the interquartile range $R_{f}$ is defined as $Q_{3}-Q_{1}$, and $k_{f}$ is a constant. After assigning $k_{f}$, we can estimate jump intensity, mean of jump size, and variance of jump size for the jump process. \par

We then estimate the MS process using maximum likelihood estimation (MLE) and the PEA process using the generalized method of moments (GMM). Following the box-plot method, we decompose the full sample into two subsamples, the diffusion subsample within the range $(Q_{1}-k_{f}R_{f}, Q_{3}+k_{f}R_{f})$ and the jump subsample outside the range $(Q_{1}-k_{f}R_{f}, Q_{3}+k_{f}R_{f})$. Based on the diffusion subsample, we estimate the MS process by the method provided in \cite{RN903}; see  Appendix \ref{MSrelerr} for the details. Based on the jump subsample, we estimate the PEA process by GMM provided in \cite{RN373}; see  Appendix \ref{PEArelerr} for the details.
The parameters for the PEA are duration $\Delta=0.02$; attenuating factor $\beta=550$;
proportional coefficient $b=4.45$.
For the MS process, the state space is
$\sigma_{k}^{2} \in \{0.0059,0.0151,0.0332,0.0577\}$, i.e, four states
with corresponding transition probability matrix
\vspace*{-8pt}
$$
P=\left[
  \begin{array}{cccc}
     0.0000   & 0.9946   & 0.0000  & 0.0054 \\
     0.2679  & 0.6506   & 0.0815  & 0.0000 \\
     0.0479  & 0.0102  & 0.9403  & 0.0016 \\
     0.0000  & 0.0062  & 0.0000  & 0.9938 \\
  \end{array}
  \right].
  $$

Next we discuss the model calibration procedure. The proposed model consists of three parts: the PEA process captures the correlation for model drivers, while the MS process and jump process capture the volatility of underlying asset. For illustrative purposes, in this example we fix the parameters for the PEA and MS processes estimated from historical data, and solve for the optimal parameters for the jump process by calibrating to the option market prices.\par

\begin{figure}[tb]
  \centering
  \includegraphics[height=8.2cm,width=11.0cm]{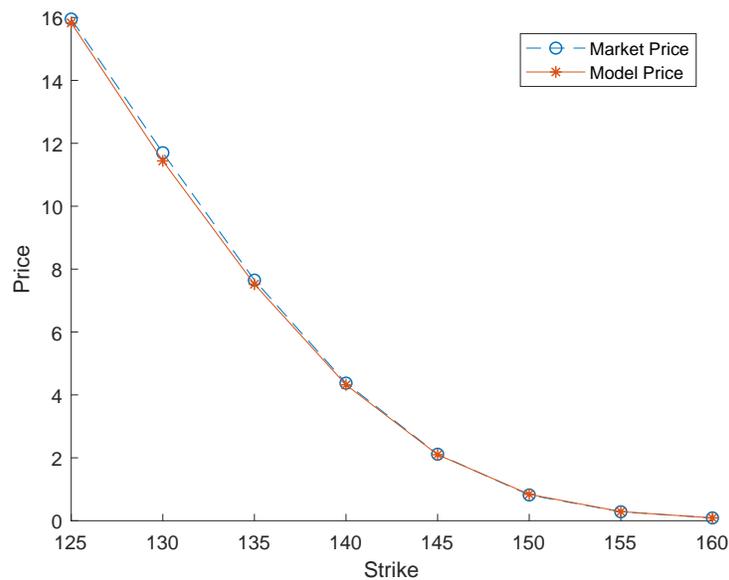}\\
  \caption{Comparison for Option Price under Market and Model}\label{figure53}
\vspace*{-9pt}
\end{figure}

Specifically, we calibrate our model to IBM call options on July 1, 2019 with maturity $T=1.5$ months. The risk-free rate $r=2.36\%$ is determined by US Dollar LIBOR rates using two maturities, 1 month and 2 months, by linear interpolation to match option maturity. Similar to Cai and Kou (2011), we minimize the objective function $\sum\limits_{i=1}^{N}(\tilde{C}_{i}(\pi)-C_{i})^{2}/C_{i}^{2}$ over the set of varying parameters $\pi=(\lambda,\mu,\sigma^{2})$ of the jump process, where $\tilde{C}_{i}(\pi)$ and $C_{i}$ represent the calibrated price and the market price for the $i$th option, respectively. To solve the optimization problem, a random search algorithm gave the final optimal solution: $\lambda=4.40$, $\mu=-0.0572$, $\sigma^{2}=0.0029$. Table \ref{table10} presents a comparison of model and market prices, where the last column shows relative biases for option price, and Figure \ref{figure53} indicates that the calibration to option prices is quite good.\par

\begin{table}[tb]
  \centering
  \caption{Call Option Prices}
    \begin{tabular}{c|ccc|c|c}
    \hline
  & \multicolumn{3}{c|}{Market} & Model &  \\[-9pt]
Strike & Bid & Ask & Mid-Price &  Price & Bias  \\
    \hline
    125 & 15.05 & 16.85 &15.95  &15.83 & $-0.75\%$  \\
    130 & 11.60 & 11.80 &11.70  &11.44 & $-2.22\%$  \\
    135 & 7.60 & 7.70 &7.65  &7.52 & $-1.70\%$  \\
    140 & 4.30 & 4.45 & 4.375  &4.32 & $-1.26\%$  \\
    145 & 2.07 & 2.16 & 2.115 &2.10 & $-0.71\%$  \\
    150 & 0.80 & 0.84 & 0.82  &0.84 & $2.44\%$  \\
    155 & 0.28 & 0.29 & 0.285  &0.29 & $1.75\%$  \\
    160 & 0.08 & 0.10 & 0.09 &0.09 & $0.00\%$  \\
    \hline
    \end{tabular}
  \label{table10}
\end{table}

\vspace*{-2pt}
\section{Conclusions}
\vspace*{-6pt}
In the paper, we propose the MS-SVCJ model to better capture volatility clustering. Under the proposed model, we derive an analytical solution for the price of European options. Due to the general nature of the model, we can apply the analytical solution to some special cases, such as the MS-SVJ model with general jump size distribution or the MS-SV model. An analytical solution for the MS-SV model avoids solving ordinary differential equations using numerical inverse Fourier transform methods. We also consider an approximation approach to price American-style options by leveraging our analytical solution for European options. To efficiently compute option prices, we propose the RR algorithm to derive the probability distribution of AIV, analyze its computational complexity, and verify its effectiveness numerically. \par

The empirical case study discussed in Section \ref{subs53} uses asset prices to estimate the MS and PEA model parameters, while calibrating the jump process parameters by market option prices. In practice, market data on actual option prices can be used to calibrate all of the model parameters of any option pricing model. Although not the focus of this work, a complete calibration procedure based on only market option prices would make our algorithm more relevant to practitioners. We briefly suggest one possible approach, which adopts a two-stage calibration procedure, cf. \cite{RN1140}, \cite{RN1150}, \cite{RN1148}, \cite{RN1149}; however, determining a good procedure is definitely a critical need for further research.

The two-stage calibration procedure calibrates the MS process and jump process (the compound Poisson process and the PEA process) separately. Specifically, at the first stage, we calibrate the MS-SV model to market option prices. The approach in \cite{RN338} appears to be well suited to our MS-SV model, since it models volatility following a MS process. In this approach, they first determine a base model reflecting one's prior information on market, then adjust the base model to fit option prices.\par

At the second stage, we can calibrate CJ component of the MS-SVCJ model with the calibrated MS-SV process at the first stage. The calibration is to determine the compound Poisson process and the PEA process by minimizing the sum of an in-sample quadratic pricing error and a convex penalization term. In practice, the key is to select the convex penalization, which consists of two terms, one due to the compound Poisson process and the other due to the PEA process.  \cite{RN1142} used relative entropy (or Kullback-Leibler divergence) from a prior distribution as a convex penalization term. For the PEA process, since parameters for the PEA process  form a Hilbert space, a quadratic function (or Tikhonov regularization) is appropriate and applied.\par

The methodology developed here should also be applicable in other contexts beyond option pricing, e.g., variance swap pricing, which depends highly on the AIV. Other open problems for future research include hedging under the proposed model, as well as extensions to more general models, e.g., models with general jump size distribution.\par

%
%
%


\begin{APPENDICES}

\section{Proof of Proposition \ref{pro2}}\label{numdis}
We denote $l_{k}$ as the number of states $\{u_{k}\}$ of sample paths of the MS process from step $1$ to step $L-1$, $k\in\{1,2,\ldots,m\}$. Thus, for each sample path of the MS process, we can get a tuple $(l_{1},l_{2},\ldots,l_{m})$. According to the definition of the weight $|\omega|$ of a sample path of the MS process, since $\sigma_{i}$, $i\in\{1,2,\cdots,L-1\}$, take value in state set $\{u_{1},\cdots,u_{m}\}$, after merging the same states in sample path, the weight can be rewritten as
$
   |\omega| = \frac{\sigma_{0}^{2}+(l_{1}u_{1}^{2}+\ldots+l_{m}u_{m}^{2})}{L}
$
where $\sigma_{0}$ is the initial state of the MS process and is pre-determined.
This leads to a surjection from tuple $(l_{1},l_{2},\ldots,l_{m})$ to weight $|\omega|$ rather than an injection due to the possibility of the different tuples generating the same weights. Thus, the number of distinct values for $v$ is less than the number of distinct tuples, which satisfies the equation
$
    L-1=l_{1}+\ldots+l_{m},
$
for which the solution is a combinatorial problem given by $\binom{L-1+m-1}{m-1}$, which is the number of ways to select $m-1$ distinct values from $\{1,2,3,\ldots,L-1+m-1\}$. Hence, we have $|\Psi| \leq \binom{L+m-2}{m-1}$. \Halmos
\endproof

\vspace*{-6pt}
\section{Proof of Lemma \ref{theo1}}\label{lemma}
According to the arbitrage-free pricing theory, the European option price is the expectation of the terminal payoff under the risk-neutral probability measure. At maturity $T$, the payoff of a European call option is $(S_{T}-K)^{+}$, so the European call option price is given by
  \begin{equation*}
  \begin{split}
  C&=e^{-rT}\mathbb{E}(S_{T}-K)^{+}\\
  &=\sum_{n=0}^{+\infty}p(N_{T}=n)e^{-rT}\mathbb{E}(S_{T}(n)-K)^{+}\\
  &=\sum_{n=0}^{+\infty}p(N_{T}=n)\sum_{\omega \in \Omega}p(\omega)e^{-rT}\mathbb{E}(S_{T}(n,\omega)-K)^{+}\\
  &=\sum_{n=0}^{+\infty}p(N_{T}=n)\sum_{\omega \in \Omega}p(\omega)\mathbb{E}_{\mathscr{J}_{n}}(e^{-rT}\mathbb{E}(S_{T}(\mathscr{J}_{n},\omega)-K)^{+})\\
  \end{split}
  \end{equation*}
  where $N_{T}$ is the number of jumps in the asset price up to $T$; $\mathscr{J}_{n}:=(J_{1},\cdots,J_{n})$ is an $n$-dimensional random vector of $n$ jump sizes; $\Omega$ is the sample path space of $\{\sigma_{k}\}$; $S_{T}(\mathscr{J}_{n},\omega)$ is the asset price at the maturity $T$ given $\mathscr{J}_{n}$ and $\omega$.  \Halmos
 \endproof

 \section{Details of Derivation for $W(\mathscr{J}_{n},\omega)$}\label{app}
 Here, we consider the conditional probability distribution of the asset price at $T$, given $n$ jumps $\mathscr{J}_{n}:=(J_{1},J_{2},...,J_{n})$ and the sample path $\omega$ of $\{\sigma_{k}\}$. Correspondingly, the asset price $S_{T}(\mathscr{J}_{n},\omega)$ is the solution to the following stochastic differential equation:
 \begin{equation*}
   \frac{dS_{t}}{S_{t_{-}}}=\left(r-\lambda \zeta\right)dt+\hat{\sigma}_{t}dB_{t}+\left(J_{t}-1\right)dN_{t}
 \end{equation*}
 where the volatility process $\{\hat{\sigma}_{t}\}$ is a deterministic function of time $t$.\par
 In terms of the analysis in \cite{RN361}, the solution to above equation is given by
 \begin{equation*}
   S_{T}(\mathscr{J}_{n},\omega)=S_{0} \prod\limits_{i=1}^{n}J_{i} e^{(r-\lambda\zeta-\frac{\mathscr{V}(\mathscr{J}_{n},\omega)}{2})T+\sqrt{\mathscr{V}(\mathscr{J}_{n},\omega)}B_{T}}
 \end{equation*}\par
 where $\mathscr{V}(\mathscr{J}_{n},\omega)=\frac{1}{T}\int_{0}^{T}\hat{\sigma}_{t}^{2}(\mathscr{J}_{n},\omega)dt$.
 Hence, we have
 \begin{equation*}
  W(\mathscr{J}_{n},\omega)=\ln\frac{S_{T}(\mathscr{J}_{n},\omega)}{S_{0} \prod\limits_{i=1}^{n}J_{i}}\Bigg|(\mathscr{J}_{n},\omega)\sim \mathcal{N}((r-\lambda\zeta-\frac{\mathscr{V}(\mathscr{J}_{n},\omega)}{2})T, \mathscr{V}(\mathscr{J}_{n},\omega)T).
  \Halmos
\end{equation*}

 \section{Proof of Theorem $\ref{the1}$}\label{tho}
Substituting Equation $(\ref{equa1})$ into Lemma $\ref{theo1}$, the European call option price is:
  \begin{equation*}
    \begin{aligned}
      C&=\sum_{n=0}^{+\infty}p(N_{T}=n)\sum_{\omega \in \Omega}p(\omega)\mathbb{E}_{\mathscr{J}_{n}}(e^{-rT}\mathbb{E}(S_{T}(\mathscr{J}_{n},\omega)-K)^{+}) \\
       &=\sum_{n=0}^{+\infty}p(N_{T}=n)\sum_{\omega \in \Omega}p(\omega)\mathbb{E}_{\mathscr{J}_{n}}(\mathbb{BS}(S_{0}e^{-\lambda\zeta T+\sum\limits_{i=1}^{n}\ln(J_{i})},|\omega|+\hat{b}\sum\limits_{i=1}^{n}\ln^{2}(J_{i}),r,T,K))\\
       &=\sum_{n=0}^{+\infty}p(N_{T}=n)\sum_{\omega \in \Omega}p(\omega)\mathbb{E}_{\Xi_{n}}(\mathbb{BS}(S_{0}e^{-\lambda\zeta T+X_{n}},|\omega|+\hat{b}Y_{n},r,T,K))\\
       &=\sum_{n=0}^{+\infty}p(N_{T}=n)\sum_{\omega \in \Omega}p(\omega)C_{n}(|\omega|)
       =\sum_{n=0}^{+\infty}p(N_{T}=n)\sum_{v \in \Psi}\sum_{\omega \in \Omega: |\omega|= v}p(\omega)C_{n}(|\omega|) \\
       &=\sum_{n=0}^{+\infty}p(N_{T}=n)\sum_{v \in \Psi}p_{V}(v)C_{n}(v),\\
    \end{aligned}
  \end{equation*}
  where $C_{n}(Z) = \mathbb{E}_{\Xi_{n}}(\mathbb{BS}(S_{0}e^{-\lambda\zeta T+X_{n}},Z+\hat{b}Y_{n},r,T,K))$, $\Xi_{n}:=(X_{n},Y_{n}):=(\sum\limits_{i=1}^{n}\ln(J_{i}),\sum\limits_{i=1}^{n}\ln^{2}(J_{i}))$. \Halmos
\endproof

\section{Proof of Proposition \ref{pro3}}\label{apdf}
For notational convenience, we denote:
\vspace*{-6pt}
\begin{equation}\label{equp21}
  \begin{aligned}
    x_{i}&=\ln(J_{i}) \\
    (x,y)&=(\sum\limits_{i=1}^{n}x_{i},\sum\limits_{i=1}^{n}x_{i}^{2}) \\
  \end{aligned}
\end{equation}
where $n$ is the number of jumps during $[0,T]$ and the jump size distribution $\ln(J_{i})\stackrel{\text{i.i.d}}{\sim}\mathcal{N}\left(\mu,\varepsilon^{2}\right)$. When $n=1$, the proof is obvious and omitted.\par

Next, we suppose $n\geq2$. Before determining the joint probability density $g(x,y)$, we determine the support set of the bivariate random variable $(x,y)$, $D=\{(x,y):g(x,y)> 0\}$. By the Cauchy-Schwarz inequality, we have:
\vspace*{-6pt}
  \begin{equation}\label{equp22}
     \left(x_{1}^{2}+x_{2}^{2}+\ldots+x_{n}^{2}\right)\left(1^2+1^2+\ldots+1^2\right)\geq \left(x_{1}+x_{2}+\ldots+x_{n}\right)^{2},
\end{equation}
so substituting the definition of $(x,y)$ in Equation $(\ref{equp21})$ into Equation $(\ref{equp22})$, we have
$ y \geq \frac{x^{2}}{n}$.
Thus, the support set $D=\{(x,y):-\infty<x<+\infty, \frac{x^{2}}{n}\leq y<+\infty\}$,
on which we derive the joint probability density $g(x,y)$. We have:
  \begin{equation*}
  \begin{aligned}
        y&=\sum_{i=1}^{n}x_{i}^{2} =\sum_{i=1}^{n}\left(\left(x_{i}-\overline{x}\right)^{2}+2x_{i}\overline{x}-\overline{x}^{2}\right)
    =\sum_{i=1}^{n}\left(x_{i}-\overline{x}\right)^{2}+2\overline{x}\sum_{i=1}^{n}x_{i}-n\overline{x}^{2} \\
    &=\left(n-1\right)s^{2}+2\frac{x}{n}x-n\left(\frac{x}{n}\right)^{2}
    =\left(n-1\right)s^{2}+\frac{x^2}{n}\\
    \Rightarrow \frac{y}{\varepsilon^{2}}&=\frac{\left(n-1\right)s^{2}}{\varepsilon^{2}}+\frac{x^{2}}{n\varepsilon^{2}}
\mbox{~ where~} \overline{x}=\frac{1}{n}\sum\limits_{i=1}^{n}x_{i}, s^{2}=\frac{1}{n-1}\sum\limits_{i=1}^{n}\left(x_{i}-\overline{x}\right)^{2}.
  \end{aligned}
  \end{equation*}

  According to Theorem 5.3.1 in \cite{RN744}, $\frac{\left(n-1\right)s^{2}}{\varepsilon^{2}}$ and $x$ are mutually independent with probability distributions
$\chi^{2}\left(n-1\right)$ and
$\mathcal{N}\left(n\mu, n\varepsilon^{2}\right)$, respectively.
By conditional probability, we decompose $g\left(x,y\right)=g\left(x\right)g\left(y \mid x\right)$, where
  \begin{equation*}
    \begin{aligned}
       g(x)&=\frac{1}{\sqrt{2\pi}\sqrt{n\varepsilon^{2}}}e^{-\frac{\left(x-n\mu\right)^{2}}{2n\varepsilon^{2}}}\\
       g(y|x) &=\frac{1}{\Gamma\left(\frac{n-1}{2}\right)2^{\frac{n-1}{2}}}\left(\frac{y-\frac{x^2}{n}}{\varepsilon^2}\right)^{\frac{n-1}{2}-1}e^{-\frac{y-\frac{x^2}{n}}{2\varepsilon^2}}\frac{1}{\varepsilon^2},
     \end{aligned}
  \end{equation*}
which leads to the desired result.  \Halmos
\endproof

\section{Proof of Corollary $\ref{cor1}$}\label{acor1}
  When $f(J_{i},t,t_{i})=0$, we have $\hat{b}=0$ and $C_{n}(Z) = \mathbb{E}_{X_{n}}(\mathbb{BS}(S_{0}e^{-\lambda\zeta T+X_{n}},Z,r,T,K))$. \par
Since $C_{n}(Z)$ does not require the probability distribution of jump size $J_{t}$, the expression holds for the jump size $J_{t}$ following a general distribution, so the European call option price is
\begin{equation*}
\begin{aligned}
C&=\sum_{n=0}^{+\infty}p(N_{T}=n)\sum_{v \in \Psi}p_{V}(v)C_{n}(v) \\
  &=\sum_{v \in \Psi}p_{V}(v)\sum_{n=0}^{+\infty}p(N_{T}=n)\mathbb{E}_{X_{n}}(\mathbb{BS}(S_{0}e^{-\lambda\zeta T+X_{n}},v,r,T,K))
  =\sum_{v \in \Psi}p_{V}(v)C_{jd}(v)
\end{aligned}
\end{equation*}
where $C_{jd}(v)=\sum\limits_{n=0}^{+\infty}p(N_{T}=n)\mathbb{E}_{X_{n}}(\mathbb{BS}(S_{0}e^{-\lambda\zeta T+X_{n}},v,r,T,K))$ is the European call option price under the jump-diffusion model. \Halmos
\endproof

\section{Complexity of RR Algorithm}\label{comrsp}
To prove the complexity of RR algorithm, we first provide the following lemma.
\begin{lemma}\label{pro1}
  The number of distinct triples $[x,l,\sigma_{l}]$ at step $l$ is less than $m\binom{l+m-2}{m-1}$.
\end{lemma}
\proof{Proof.}
At step $l$,  the number of distinct triples $[x,l,\sigma_{l}]$ can be decomposed into the product of the number of distinct values of $x$ and the number of distinct values of $\sigma_{l}$, which is a combinatorial problem. Obviously, the number of distinct values of $\sigma_{l}$ is $m$. To determine the number of distinct values of $x$, we denote $l_{k}$ as the number of states $\{u_{k}\}$ of subsample paths of the MS process from step $1$ to step $l-1$, $k\in\{1,2,\ldots,m\}$. Thus, for each subsample path of the MS process, we can get a tuple $(l_{1},l_{2},\ldots,l_{m})$. Similar to Proposition \ref{pro2}, we have:
\begin{equation*}
  \begin{aligned}
    |\omega_{l}|&=\frac{\sigma_{0}^{2}+(l_{1}u_{1}^{2}+\ldots+l_{m}u_{m}^{2})}{l} \\
    l-1&=l_{1}+\ldots+l_{m}
  \end{aligned}
\end{equation*}
where $\sigma_{0}$ is the initial state of the MS process and is pre-determined.\par

According to Proposition \ref{pro2}, the number of distinct values of $x$ is less than $\binom{l-1+m-1}{m-1}$. Hence, the number of distinct triples $[x,l,\sigma_{l}]$ at step $l$ is less than $m\binom{l+m-2}{m-1}$.\Halmos
\endproof

\begin{proposition}
  The total number of distinct triples $[x,l,\sigma_{l}]$ from step $1$ to step $L$ is less than $m\binom{L+m-1}{m}$, hence the complexity of the RR algorithm is $O\left(L^{m}\right)$.
\end{proposition}
\proof{Proof.}
  The total number of distinct triples $[x,l,\sigma_{l}]$ from step $1$ to step $L$ is a summation of the number of distinct triples $[x,l,\sigma_{l}]$ at every step $1\leq l\leq L$. According to Lemma \ref{pro1} providing an upper bound for the number of distinct triples $[x,l,\sigma_{l}]$ at step $l$, carrying out a summation for $L$ steps, we will provide an upper bound for the total number of distinct triples $[x,l,\sigma_{l}]$ from step $1$ to step $L$,
\begin{equation*}
\begin{split}
\sum_{i=1}^{L}m\binom{i-1+m-1}{m-1}=m\sum_{i=0}^{L-1}\binom{i+m-1}{m-1}&=m\left[\binom{m-1}{m-1}+\sum_{i=1}^{L-1}\binom{i+m-1}{m-1}\right]\\
&=m\left[\binom{m}{m}+\sum_{i=1}^{L-1}\binom{i+m-1}{m-1}\right]\\
&=m\left[\binom{m+1}{m}+\sum_{i=2}^{L-1}\binom{i+m-1}{m-1}\right]\\
&\ldots \\
&=m\binom{L-1+m}{m}.
\end{split}
\end{equation*}
Hence, the total number of distinct triples $[x,l,\sigma_{l}]$ from step $1$ to step $L$ is less than $m\binom{L+m-1}{m}$. \par
In addition, since $L\gg m$ in settings of practical interest, we have:
$$
m\binom{L-1+m}{m} = m \frac{(L-1+m)!}{m!(L-1)!}
= \frac{L+m-1}{m-1} \cdot \frac{L+m-2}{m-2} \cdots \frac{L+1}{1} \cdot L
~=~O(L^{m}). \Halmos
$$
\endproof

\clearpage
\section{MATLAB Code for RR Algorithm}\label{rrsp}
\begin{lstlisting}
function [leftvariance,leftprob]=AveStdTest5(iniprob,variance,matrix,n)
% iniprob: the initial state of the MS process, e.g., [0 1 0 0];
% variance: the state space of variance in ascending order, e.g., [0.02 0.04 0.06 0.08]
% matrix: the transition matrix P;
% n:  the total number of time steps L;
tic;

leftvariance=transpose(variance)+dot(iniprob,variance);
leftprob=transpose(iniprob*matrix);
nstep=n-1;
nstate=size(matrix,1);
for i=2:nstep
    leftvariance=leftvariance+variance;
    transmat=repmat(matrix,size(leftvariance,1)/nstate,1);
    leftprob=leftprob.*transmat;

    group=findgroups(round(leftvariance(:,1),10));
    leftprobmake=zeros(max(group),nstate);leftvariancemake=zeros(max(group),nstate);
    for j=1:nstate
        leftprobmake(:,j)=accumarray(group,leftprob(:,j),[],@sum);
        leftvariancemake(:,j)=accumarray(group,leftvariance(:,j),[],@min);
    end

    leftprob=transpose(leftprobmake);
    leftvariance=transpose(leftvariancemake);
    leftprob=leftprob(:);
    leftvariance=leftvariance(:);
end

group=findgroups(round(leftvariance(:,1),10));
leftprobmake=accumarray(group,leftprob,[],@sum);
leftvariancemake=accumarray(group,leftvariance,[],@min);

leftvariance=leftvariancemake/n;
leftprob=leftprobmake;
toc;
end
\end{lstlisting}

\clearpage

\section{Impact of Assumption on Jump Time}\label{relerr}
We discuss the impact of the assumption that all jumps in the small interval occur at the beginning of the interval on the asset price and AIV. Since the cumulative impact on asset price does not relate to the actual times of the jumps, the asset price at $T$ does not change under the assumption.
In what follows, considering the probability of a jump, we analyze the expectation bias ($EB$) caused by the assumption on AIV.\par
First, we denote the jump probability and the expectation bias as $P_{l}$ and $EB_{l}$, respectively, when there are $l$ jumps up to maturity $T$, given by
\begin{equation*}
\begin{aligned}
  EB & =\sum_{l=1}^{+\infty}P_{l}*EB_{l}, \\
  P_{l} & =\frac{(\lambda T)^{l}}{l!}e^{-\lambda T}.\\
\end{aligned}
\end{equation*}\par
Second, since the sample path of the MS process and the jump during the interval $[0,T-\Delta]$ do not cause the bias, we only investigate the bias caused by a jump during the interval $[T-\Delta,T]$. Given $l$ jumps up to maturity $T$, we denote the conditional jump probability and the expectation bias as $P_{j}^{l}$ and $EB_{j}^{l}$, respectively, for $1 \leq j \leq l$ jumps during the interval $[T-\Delta,T]$, given by
\begin{equation*}
\begin{aligned}
  EB_{l} & = \sum_{j=1}^{l}P_{j}^{l}*EB_{j}^{l}\\
  P_{j}^{l} &=\frac{\frac{(\lambda \Delta)^{j}}{j!}e^{-\lambda \Delta}*\frac{(\lambda (T-\Delta))^{l-j}}{(l-j)!}e^{-\lambda (T-\Delta)}}{P_{l}}\\
\end{aligned}
\end{equation*}
where
\begin{equation*}
  \begin{aligned}
P_{j}^{l}& =p(N_{(T-\Delta,T)}=j\mid N_{(0,T)}=l)
  =\frac{p(N_{(T-\Delta,T)}=j,N_{(0,T)}=l)}{p(N_{(0,T)}=l)}\\
  &=\frac{p(N_{(T-\Delta,T)}=j,N_{(0,T-\Delta)}=l-j)}{p(N_{(0,T)}=l)}
  =\frac{p(N_{(T-\Delta,T)}=j)p(N_{(0,T-\Delta)}=l-j)}{p(N_{(0,T)}=l)}.
   \end{aligned}
\end{equation*}\par
Third, we derive the detailed expression for $EB_{j}^{l}$. For the $i$th$(1\leq i\leq j)$ jump $J_{i}$ at time $t_{i}\in[T-\Delta,T]$, without or with the assumption, the cumulative effects until expiration date $T$ are, respectively:
 \begin{equation*}
\begin{aligned}
   \int_{t_{i}}^{T}b\ln^{2}(J_{i})e^{-\beta (s-t_{i})}ds &=\frac{b\ln^{2}(J_{i})}{\beta}(1-e^{-\beta (T-t_{i})}), \qquad \text{Without the assumption}, \\
  \int_{T-\Delta}^{T}b\ln^{2}(J_{i})e^{-\beta(s-(T-\Delta))}ds &=\frac{b\ln^{2}(J_{i})}{\beta}(1-e^{-\beta \Delta}),  \qquad \text{With the assumption}. \\
\end{aligned}
\end{equation*}
Hence,
\begin{equation*}
  \begin{split}
EB_{j}^{l}&=\frac{1}{T}\mathbb{E}(\sum_{i=1}^{j}(\frac{b\ln^{2}(J_{i})}{\beta}(1-e^{-\lambda\Delta})-\frac{b\ln^{2}(J_{i})}{\beta}(1-e^{-\lambda(T-t_{i})})))\\
&=\frac{b\eta}{\beta T}\mathbb{E}(\sum_{i=1}^{j}(1-e^{-\beta\Delta})-(1-e^{-\beta(T-t_{i})}))\\
&=\frac{b\eta}{\beta T}\mathbb{E}(\sum_{i=1}^{j}(e^{-\beta Y_{i}}-e^{-\beta\Delta}))\\
&=\frac{jb\eta}{\beta T}(\frac{1-e^{-\beta\Delta}}{\beta\Delta}-e^{-\beta\Delta})\\
   \end{split}
\end{equation*}
where $\eta=\mathbb{E}(\ln^{2}(J_{i}))=\mu^2+\varepsilon^2$ and $T-t_{i}=Y_{i}\sim U[0,\Delta]$, where $U[0,\Delta]$ is a uniform distribution, since for the Poisson process with intensity $\lambda$, conditioned on $N_{t}=n$, the joint probability distribution of the ordered arrival times of jumps $t_{1}<t_{2}<\cdots<t_{n}$ is the same as the joint probability distribution of the order statistics $U_{(1)}<U_{(2)}<\cdots<U_{(n)}$ with $U_{i} \stackrel{\text{i.i.d.}}{\sim} U[0,t]$, $i=1,2,\cdots,n$.

Since $EB_{j}^{l}\rightarrow EB_{l}\rightarrow EB$, we have
\begin{equation*}
  EB =\sum_{l=1}^{+\infty}\sum_{j=1}^{l}\frac{(\lambda \Delta)^{j}}{j!}e^{-\lambda \Delta}*\frac{(\lambda (T-\Delta))^{l-j}}{(l-j)!}e^{-\lambda (T-\Delta)}\frac{jb\eta}{\beta T} \frac{1-(1+\beta\Delta)e^{-\beta\Delta}}{\beta\Delta}.
\end{equation*}
Taking $N_{max}=10$, $T=0.25$, $\lambda=3$, $\beta=250$, $\Delta=0.02$, $\mu=-0.025$, $\varepsilon^2=0.005$ in Table \ref{table6} as an example, $EB=2.07\times10^{-6}$. The option price $C=0.9696$ implies volatility $\sigma_{imp}=0.2475$, so the assumption increases volatility by $\sqrt{\sigma_{imp}^2}-\sqrt{\sigma_{imp}^2-EB}=4.18\times10^{-6}$, which is less than 0.002$\%$. 

\section{Estimating Parameters in the PEA Process}\label{para}
We describe estimation of the parameters of the PEA process: proportional coefficient $b$, attenuating factor $\beta$ and duration $\Delta$. For this purpose, we adopt the approach of \cite{RN373}, in which the modeling of co-jumps is similar to ours, viz., the jump in variance is also proportional to the squared jump in return and exponentially decays over time. \par
Once a jump in return occurs, the proportional coefficient $b$ determines the corresponding increment of variance. In terms of the expressions of $m_{c}$ and $m_{d}$ in \cite{RN373}, we derive the proportional coefficient $b=2$. \par
The function $f(u)$ in \cite{RN373} describes the evolving pattern of jump in variance, which corresponds to our function $f_{2}(\cdot)$. For $\beta$, through sampling points $\{(u_{i},f(u_{i}))\}_{i=1}^{n}$ from $f(u)$ in \cite{RN373} and implementing the least squares method, we estimated the attenuating factor $\beta=250$ with the goodness of fit, $R^{2}=0.77$. \par

We select the duration $\Delta=0.02$, which means once there is jump in variance, our model can cover $99.33\% \approx 1-\frac{e^{-250*0.02}}{e^{-250*0}}$ of this increment over the next 5 days. We assume a year includes 252 trading days, hence $252*0.02\approx 5$ days.

\section{Additional Empirical Results on Computational Complexity for CE and RR}\label{plot}
We confirm the theoretical computational complexity of the CE and RR algorithms with a larger number of states $m\in\{2,3,4,5,6\}$. Specifically,  the computation time as a function of the number of time steps $L$ is shown in Figure \ref{plot1}, in line with the theoretical results and numerical experiments in the main body of the main manuscript.
\vspace*{-14pt}
\begin{figure}[ht]
  \centering
\subfigure[CE with $m=2,3$]{
\begin{minipage}{0.48\linewidth}
  \centering
  \includegraphics[height=6.0cm,width=8.0cm]{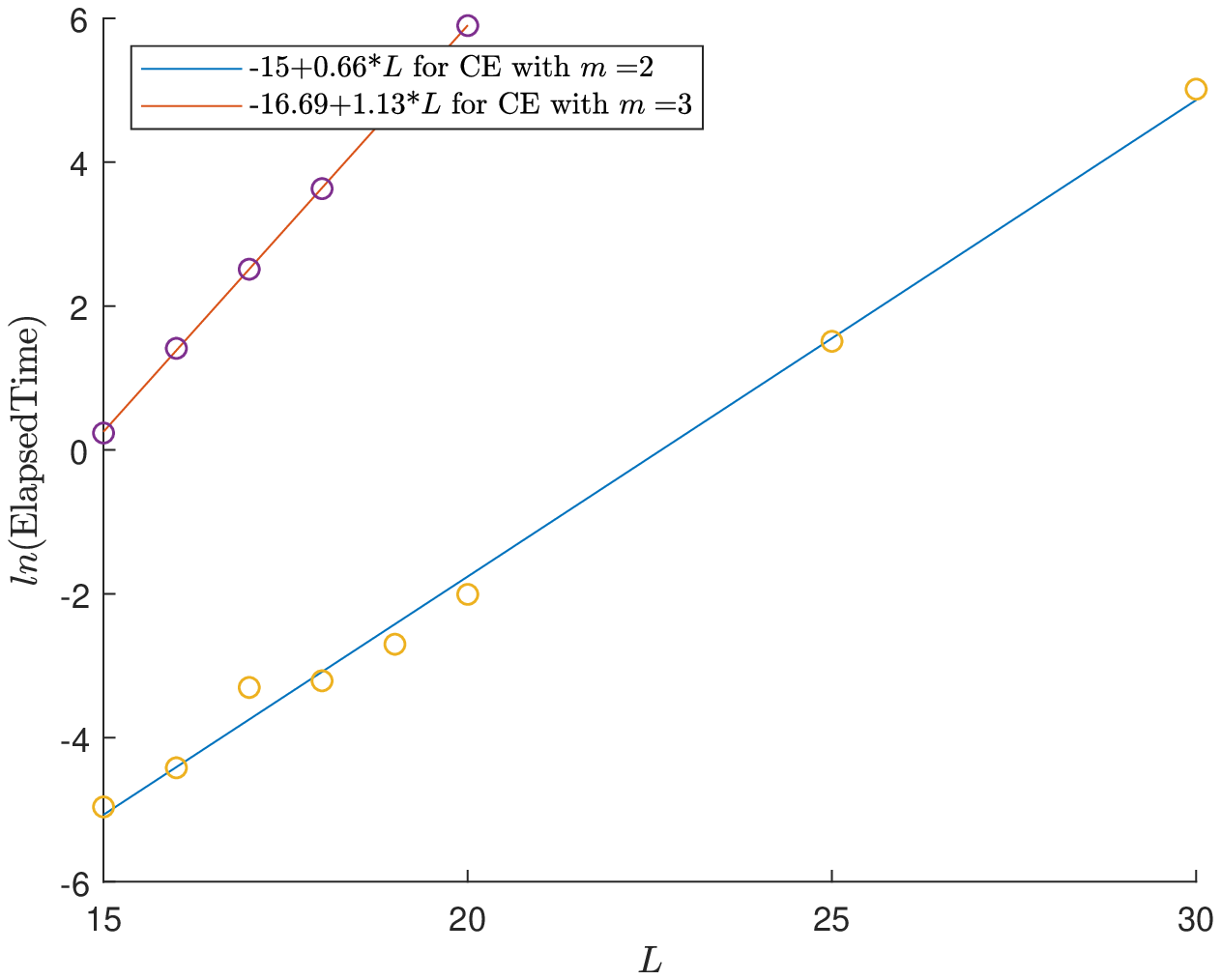}\\
\end{minipage}
}
\subfigure[RR Algorithm with $m=2,3,4,5,6$]{
\begin{minipage}{0.48\linewidth}
  \centering
\includegraphics[height=6.0cm,width=8.0cm]{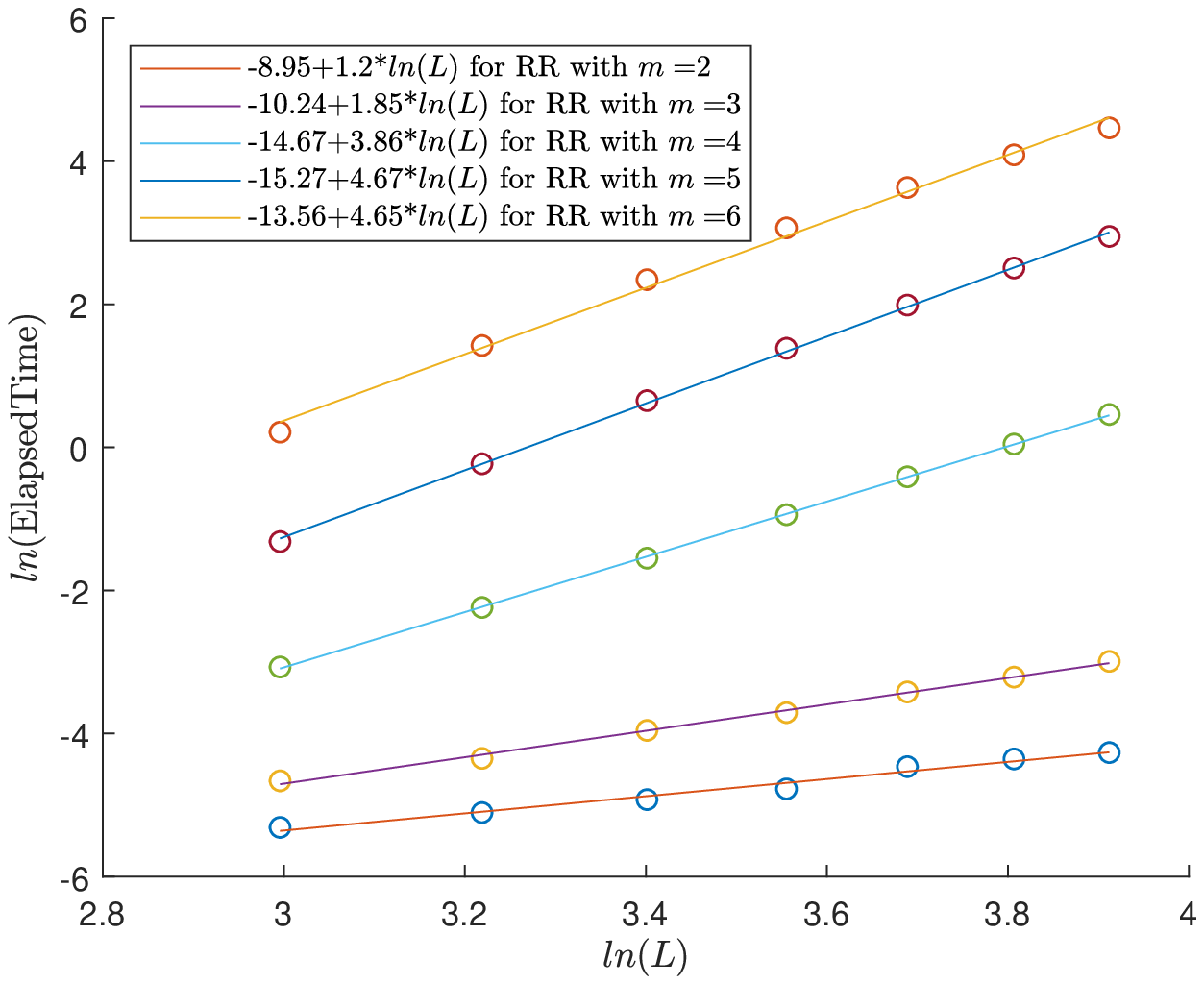}\\
\end{minipage}
}
\caption{Computation Time as a Function of the Number of Time Steps $L$ (Log Scales)}
\label{plot1}
\end{figure}

\vspace*{-14pt}

\section{Estimation of the MS Process for the Application Example in Section \ref{subs53}}\label{MSrelerr}  
Following the standard risk premia assumptions in the literature, the asset price without jumps under the objective probability measure follows geometric Brownian motion with drift  $\vartheta$ and MS stochastic volatility $\sigma_{t}$. To estimate the MS process, we consider the discrete version of the asset price described by
$$
  \frac{S_{t+a}-S_{t}}{S_{t}}=\vartheta a+\sigma_{t}\mathcal{N} \sqrt{a}  \Longrightarrow \tilde{r}^{a}_{t}=\frac{S_{t+a}-S_{t}}{\sqrt{a} S_{t}}=\vartheta \sqrt{a}+\sigma_{t}\mathcal{N},
$$
where $\mathcal{N}$ follows a standard normal distribution.
Using 
this result with existing MATLAB codes provided in \cite{RN903} to the diffusion subsample generating $\tilde{r}^{a}_{t}$, the estimated parameters for MS process are easily obtained. \Halmos

\section{Estimation of the PEA Process for the Application Example in Section \ref{subs53}}\label{PEArelerr} 
Given the path of the MS process $\{\sigma_{t}\}$, we derive closed-form expressions for variance, skewness, kurtosis of asset log-return:
\begin{equation}\label{equa30}
  \begin{aligned}
    \mathbb{E}(r^{a}_{t}-\mathbb{E}(r^{a}_{t}))^{2}& =a\sigma_{t}^{2}+a(1+\frac{b}{\delta})M_{2} \\
    \mathbb{E}(r^{a}_{t}-\mathbb{E}(r^{a}_{t}))^{3}& = (a+\frac{3b}{\delta^2}(\delta a-1+e^{-\delta a}))M_{3} \\
    \mathbb{E}(r^{a}_{t}-\mathbb{E}(r^{a}_{t}))^{4}& = 3(a\sigma_{t}^{2})^2+6a^2\sigma_{t}^{2} (1+\frac{b}{\delta})M_{2}+ \\ &(a+\frac{6b}{\delta^2}(\delta a-1+e^{-\delta a}))M_{4}+(\frac{3a^2b(b+2)}{\delta^2}+3a)M_{2}^{2} \\
  \end{aligned}
\end{equation}
where $M_{i}=\lambda m_{i}$, and $m_{i}$, $i=2,3,4$ are the $i$th moments of the log-jump distribution.

Specifically, given the path of the MS process $\{\sigma_{t}\}$, our original model in Equation $(\ref{original})$ has similar probability characteristics as the model in \cite{RN373}, and Equation $(\ref{equa30})$ can be justified by Theorem 1 of \cite{RN373}. Applying the generalized method of moments (GMM) estimation to the jump subsample, the estimated parameters for the PEA process are easily obtained. 

\end{APPENDICES}

\vspace{-10pt}
\ACKNOWLEDGMENT{\baselineskip14pt
Fu gratefully acknowledges financial support from the U.S. National Science Foundation [Grant CMMI-1434419]. Li gratefully acknowledges financial support from the Natural Science Foundation of China [Grant 71671094], and the Fundamental Research Funds for the Central Universities [Grants 63185019 and 63172308]. The views and opinions expressed in this article are solely the authors own and do not reflect the business and positions of R. Wu's affiliation.}
\vspace{-6pt}

\baselineskip18.8pt

\bibliographystyle{informs2014}
\bibliography{reflib} 

\end{document}